\documentclass[aps,pre,showpacs]{revtex4}
\usepackage{graphicx}
\usepackage{dcolumn}
\usepackage{bm}

\begin{document}

\title{Photonic band gaps in materials with triply periodic surfaces\\
and related tubular structures}
\author{
K. Michielsen\footnote{E-mail: kristel@phys.rug.nl} and
J.S. Kole\footnote{E-mail: j.s.kole@phys.rug.nl}
}
\affiliation{%
Applied Physics - Computational Physics\footnote{http://www.compphys.org},
Materials Science Centre\\
University of Groningen, Nijenborgh 4\\
NL-9747 AG Groningen, The Netherlands
}
\date{\today}

\begin{abstract}
We calculate the photonic band gap of triply periodic
bicontinuous cubic structures and of tubular structures
constructed from the skeletal graphs of triply periodic minimal surfaces.
The effect of the symmetry and topology
of the periodic dielectric structures on the existence and the
characteristics of the gaps is discussed.
We find that the C(I$_2$-Y$^{**}$) structure with $Ia{\bar 3}d$ symmetry,
a symmetry which is often seen in experimentally realized bicontinuous
structures, has a photonic band gap with interesting characteristics.
For a dielectric contrast of 11.9 the largest gap
is approximately 20\% for a volume fraction of the high dielectric material of 25\%.
The midgap frequency is a factor of 1.5 higher than the
one for the (tubular) D and G structures.
\end{abstract}

\pacs{42.70.QS, 41.20.Jb, 02.40.-k}

\maketitle

\def\br{{\mathbf{r}}}

\section{Introduction}
Photonic crystals are periodic dielectric composite structures that forbid
propagation of electromagnetic waves, in any direction and for any polarization,
for a certain frequency range, called the photonic band gap (PBG).
PBGs are characterized by two quantities: the width of the bandgap and
the midgap frequency. The characteristics of PBGs in periodic
dielectric structures depend on the dielectric contrast between the composites,
the symmetry and topology of the structure and the filling factor,
the ratio between the volume occupied by each dielectric with respect to
the total volume of the composite.
 
Many applications of photonic crystals are in the visible (400-700 nm)
or near-infrared wavelength (700-1300 nm) range~\cite{PBG1,PBG2,PBG3,PBG4,PBG5,PBG6}.
Finding three-dimensional (3D) photonic structures with a large
absolute band gap in the visual or near-infrared frequency range
and suitable for large-scale production is therefore highly desirable.
Because of the sub-micrometer resolution required in the production technology,
manufacturing these 3D photonic crystals is a great challenge.
Techniques involved in the
manufacturing process of the PBG materials include
microfabrication, microlithography, self-assembly methods, weaving of
dielectric fibers and the glancing angle deposition technique~\cite{GLAD}.

In the two last decades several 3D photonic crystal architectures have been proposed and/or
produced. Amongst them are the criss-crossing pore structures (inverse diamond
lattice of overlapping air spheres in a high dielectric background)~\cite{Ho0,Yablonovitch0},
the woodpile structure~\cite{Woodpile1,Woodpile2}, inverse opal structures~\cite{Miguez,Wijnhoven,Blanco},
circular spiral structures~\cite{Chutinan0},
structures from woven dielectric fibers~\cite{Tsai}, square spiral structures~\cite{Toader0,Toader1},
self-assemblies of metal nanospheres~\cite{Moroz1,Wang1}
and bicontinuous cubic microstructures~\cite{Minsurf3,Minsurf4}.
In self-assembling systems many cubic phases with various symmetries and topologies have been discovered,
see for example Refs.~\cite{Strom,Funari,Alexandridis,Sakamoto}.
However, the list of the experimentally identified cubic phases may be incomplete because
the analysis of the scattering patterns of these complex structures can be very difficult~\cite{Holyst1,Holyst2}.
In self-assembling systems the gyroid structure with $Ia{\bar 3}d$ symmetry is observed quite often.
Mesoporous solids (silicas) exhibiting these gyroid morphologies have also been obtained~\cite{Monnier,Chan,Lee}.
It was suggested that these structures have potential as PBG materials~\cite{Chan}. Recently, however,
it has been shown theoretically that these double gyroid structures do not exhibit a PBG~\cite{Minsurf4}.
The more simple cubic structures such as the P ($Pm{\bar 3}m$), D ($Fd{\bar 3}m$) and
G($I4_132$) structures can develop PBGs~\cite{Minsurf3,Minsurf4}.

The purpose of this paper is to calculate the band gaps for a large variety of triply periodic
bicontinuous cubic structures: The P,
D, G, C(P), C(D), Y, C(Y), $^{\pm}$Y, C($^{\pm}$Y), C(Y$^{**}$)=C(G),
S, C(S), I-WP, F-RD, I$_2$-Y$^{**}$, C(I$_2$-Y$^{**}$), and L ~\cite{Level1,Schnering} structures and the
P, D and G tubular structures.
Thereby we focus on the influence of the symmetry and topology
of the dielectric structure on the appearance of a PBG.

The paper is organized as follows. Section 2 describes how we generate the
different periodic dielectric structures and how we calculate the PBGs. In sections 3 and 4 we discuss
the results. Section 5 contains our conclusions.

\section{Theory}

In mesoscopic self-assembled bicontinuous cubic structures, interfaces separate adjacent
regions of different composition. These interfaces are triply periodic,
i.e. they have a 3D Bravais lattice,
and are often called intermaterial dividing surfaces (IMDS). IMDS can be approximated
by constant mean curvature surfaces which can be modelled
by level surfaces ~\cite{Level1}.
Single level surfaces, dividing space into two infinite, connected but disjunct regions, are defined
by~\cite{Level1,Schnering}
\begin{equation}
f(x,y,z)=\sum_{hkl}|F(hkl)|\cos (hX+kY+lZ-\alpha_{hkl})=t,
\end{equation}
where $X=2\pi x/a$, $Y=2\pi y/a$, $Z=2\pi z/a$, $(x,y,z)$ are the positions of atoms in
the crystal structure and
$a$ denotes the length of the crystallographic cell, the smallest
cube generating space by the lattice~\cite{Grosse}.
$\alpha_{hkl}$ and $|F(hkl)|$ denote the phase angle and the
structure factor amplitude, reflecting the symmetry of the structure, respectively.
The parameter $t$
determines the volume fraction of the two regions, or labyrinths.
For $t=0$, Eq. (1) defines the nodal surfaces~\cite{Schnering}. They are used as
approximations to the triply periodic minimal surfaces, surfaces
for which the mean curvature is zero at every point.
Note that $t=0$ does not always divide space in two regions of equal volume.
The labyrinth volume of the single level surfaces decreases monotonically with increasing
absolute value of $t$~\cite{Minsurf1}.
At $t_1$ and $t_2$, where $t_1<0<t_2$, the surfaces ``pinch-off'', i.e. become disconnected. They do
no longer subdivide space into two continuous subvolumes
but reduce to a lattice of closed-packed units with a given symmetry~\cite{Minsurf1,Anderson}.
The pinching off behavior, and hence the change in topology can be controlled by considering
the skeletal graphs of triply periodic surfaces generated using level set methods~\cite{website}.
After the surfaces pinched-off, they disappear completely for some $t<t_1$ or $t>t_2$.

The two regions defined by the single level surfaces are not simply connected.
They interpenetrate each other in a complicated way.
The topological complexity of the level surfaces can be
characterized by the genus $g$~\cite{Minsurf5}.
The genus counts the number of handles, i.e. the number of holes in the closed surface.
A finite surface of genus g is the topological equivalent of a sphere with g handles.
The genus of a sphere is zero, whereas the genus of a torus with one hole is one.
In what follows we give the genus for the
crystallographic cell.
The genus is computed using the integral-geometry morphological
image analysis method~\cite{Minsurf5}.
The two labyrinths may differ in shape or there may
exist symmetry operations mapping one labyrinth onto the other.
For the case of the PBG structures we consider the two regions to consist of material with
a different dielectric constant. This makes the two labyrinths always distinguishable,
even if they have the same shape.
The symmetry properties of the level surface structures are therefore described by the crystallographic space group
that does not include symmetry operations which would interchange the different dielectric regions.

The PBG properties of dielectric structures with IMDS that can be
modelled by P, D, G minimal~\cite{Minsurf3,Minsurf4} and level surfaces~\cite{Minsurf4}
have been studied previously for a dielectric contrast of 13.
In this paper we report on band gap calculations for periodic dielectric structures
defined by the single level surfaces P,
D, G, C(P), C(D), Y, C(Y), $^{\pm}$Y, C($^{\pm}$Y), C(Y$^{**}$)=C(G),
S, C(S), I-WP, F-RD, I$_2$-Y$^{**}$, C(I$_2$-Y$^{**}$), and L ~\cite{Level1,Schnering} and by the skeletal graphs of the
P, D and G minimal surfaces~\cite{website}. Since the I-WP, F-RD, I$_2$-Y$^{**}$, C(I$_2$-Y$^{**}$) and L
surfaces for $t=0$ do not divide space in two regions of equal volume, we investigate both the
direct and the inverse dielectric structures defined by these surfaces.
The cubic periodic dielectric structures
consist of $4\times 4\times 4$ unit cells of length $a=3\lambda$,
where $\lambda$ is the unit of length, and
are characterized by magnetic
permeability $\mu(x,y,z)=1$ and dielectric constant
\begin{equation}
\varepsilon(x,y,z)=\left\{\begin{array}{c}\varepsilon_s,
\quad\text{if}\quad f(x,y,z) < t\\ \varepsilon_b,
\quad\text{if}\quad f(x,y,z) \geq t \end{array}\right.,
\end{equation}
where $\varepsilon_b$ and $\varepsilon_s$ denote the dielectric
constant of the background and structure, respectively.
We choose the dielectric constant of the
structure material to be that of Si ($\varepsilon =11.9$)~\cite{Palik} and
the one of the background material to be that of
air ($\varepsilon =1$).
The volume fraction or filling ratio of the dielectric material
forming the background and structure are denoted by $\phi_b$ and
$\phi_s$, respectively. Varying the parameter $t$ changes $\phi_s$
and $\phi_b$ and hence the composition of the material.
The inverse structures are defined by
\begin{equation}
\varepsilon(x,y,z)=\left\{\begin{array}{c}\varepsilon_s,
\quad\text{if}\quad f(x,y,z) \geq t\\ \varepsilon_b,
\quad\text{if}\quad f(x,y,z) < t \end{array}\right..
\end{equation}

For the various periodic dielectric structures we obtain the band gap maps,
showing the band gap width
as a function of the volume fraction $\phi_s$, from the computation of
the density of states (DOS), or eigenvalue distribution.
The DOS is calculated using an unconditionally
stable finite difference time domain (FDTD) method to solve the time-dependent Maxwell
equations~\cite{Kole01,Kole02}. The algorithm used is the T4S2 algorithm
and the algorithm-specific parameters are $\tau=\Delta t=0.075\lambda /c$,
$\delta=0.1\lambda$, $M=8192$ (see Refs. ~\cite{Kole01,Kole02}).
The width of the bandgap is defined as the width $\Delta\omega$ of the frequency
range of the bandgap divided by the midgap frequency $\omega_0$ and is expressed in
percentages.
Angular frequencies are given in units of $2\pi c/a$ where $c$ is the speed of light in vacuum.

\section{Level surface structures: Results}
In Tables I-IV we give the equations for the direct and inverse level surface structures together
with 3D renderings of 2x2x2 unit cubes of the structures for various values of $t$.
Note that the equations for the direct and inverse structures are the same. The structures become
different through Eq.(2) and Eq.(3) for the direct and inverse structure, repectively.
The interval of $t$ values is chosen such that the endpoints approximate the
$t$ values for pinch-off of the surface and such that the range of $t$ values for which the structures show
a PBG is covered.
As can be seen from Tables I-IV the length of these
intervals can be quite different for the various level surface structures.
The corresponding volume fractions $\phi_s$ are also given.
The structures are listed in the order of increasing index of their space group~\cite{ITC}
(see also Table V) and not in order of appearance in the text.

We first consider the P, D and G level surface structures.
For the P surface a family of connected level surfaces exists
for $-1.45\leq t\leq 1.45$, corresponding to $\phi_s$
values between 0.14 and 0.86, where we calculated the volume fractions $\phi_s$
by means of the morphological image analysis method~\cite{Minsurf5}.
The D level surface family exists from a volume fraction
$\phi_s$ of 0.16 to 0.84 corresponding to $t$ values between 0 and $|t_1|=|t_2|=1.00$.
The G level surfaces pinch-off at $|t_1|=|t_2|=1.41$ which corresponds to
values of $\phi_s$ of 0.02 and 0.98.

Fig.\ref{fig:levsurf1} shows the band gap maps for the P, D and G level surface structures.
Gaps for a broad range of volume fractions are observed for the P,
D and G structures, in agreement with the results presented in Ref.~\cite{Minsurf4}.
For the P structure a gap is found for volume fractions $\phi_s$
in the range 12-33 \%, corresponding to $t$ values between -1.50 and -0.60. Note that
for $-1.50<t<-1.45$ the structure is disconnected and consists of ``spheres'' on
a simple cubic lattice.
The largest gap $\Delta\omega /\omega_0=6.8\%$ is found for a volume fraction of approximately 24\%.
The corresponding midgap frequency $\omega_0=0.442$.
A PBG is absent for $\phi_s=50\%$ ($t=0$), in agreement with Refs. ~\cite{Minsurf3,Minsurf4}.
The band gap maps for D and G level surface structures show that compared to
the P structure, the D and G structures have gaps for a much broader
range of volume fractions. The maximum value for the D and G level structures
is reached for volume fractions of
0.25 ($t=-0.60$) and 0.21 ($t=-0.90$), respectively.
For the D (G) structure $\Delta\omega /\omega_0=20.3\%$ (21.7\%) and
$\omega_0=0.541$ (0.474).
Comparison of the band gap maps in Fig.1 with the ones in
Ref.~\cite{Minsurf4} shows that the PBGs in Fig.1 close at slightly lower
filling fractions. This may be due to the fact that we have chosen a
lower dielectric contrast between the media. In general, lowering the
dielectric contrast decreases the band gap width.
In contrast to the numerical method used in Ref.~\cite{Minsurf4}
our method puts accurate bounds on the gap, even at low volume fractions.
Hence the band gap maps in Fig.\ref{fig:levsurf1} for the P and D level surface structures
close, in contrast to the ones shown in Ref.~\cite{Minsurf4}.
The maximum value for the bandgap widths, the midgap frequency and the corresponding filling
fractions are summarized in Table V.

As mentioned above the characteristics of a band gap not only depend on
the dielectric contrast and the volume fractions of the two media
but also on the symmetry and the topology of the structure.
The P, D and G level surfaces have $Pm{\bar 3}m$, $Fd{\bar 3}m$ and $I4_132$
symmetry, respectively~\cite{Schnering,Fisher}. The P surface has genus 3, the D surface has $g=9$
and the G surface has $g=5$~\cite{Minsurf5}. Thus, it is of interest to
repeat the band gap analysis for level surface structures having the
same symmetry but with different topologies.

We first consider the space group $Pm{\bar 3}m$~\cite{ITC} of the P surface to which also the
C(P) and I-WP surfaces belong~\cite{Schnering,Fisher}.
The C(P) level surfaces with $g=9$ pinch-off at $|t_1|=|t_2|=0.67$ which corresponds to
values of $\phi_s$ of 0.28 and 0.72. These volume fractions are relatively
high compared to the volume fraction at which the largest PBG for the P structure occurs.
For the I-WP surface, which has genus 7, a family of connected level surfaces exists
for $-3.00\leq t\leq 2.98$, corresponding to $\phi_s$
values between 0.10 and 0.99.
Band gap calculations for both structures show no gap
at any volume fraction for $t_1\le t\le t_2$.
Since for the I-WP surface, $t=0$ does not divide space in regions of equal volume, we also
study the inverse I-WP structure. Also for this structure we find no PBG.
Hence, for the level surface structures with $Pm{\bar 3}m$ symmetry only the structure with the
lowest genus, and hence the simplest topology, has a PBG.

A level surface which has the same symmetry as the D surface is the C(D) surface~\cite{Schnering,Fisher}.
The C(D) level surface family only exists for relatively high volume fractions:
$0.35\le\phi_s\le 0.58$ corresponding to $-0.28\le t\le 0.26$.
The C(D) surface
has genus 121. For the C(D) surface, two different values for $g$
for the crystallographic cell can be found in the literature: $g=145$ as found by
Fisher {\sl et al.}~\cite{Fisher} for example, and $g=121$ as found by
Garstecki {\sl et al.}~\cite{Garstecki2}.
This difference might be caused by the fact that Fisher {\sl et al.} have worked with minimal
surfaces, while Garstecki {\sl et al.} have used the nodal approximations, as we do in this work.
The same reasoning can be applied to explain the fact that the C(D) level surface does not
divide space in regions of equal volume for $t=0$, while the minimal C(D) surface does~\cite{Fisher}.
Our calculations indicate that the C(D) structure shows no PBG.

The C(Y$^{**}$)=C(G) surface,
also called the complementary gyroid~\cite{Schnering},
has the same symmetry as the G surface and has $g=29$.
The C(Y$^{**}$)=C(G) level surface family with $g=29$ exists from a volume fraction
$\phi_s$ of 0.29 to 0.71 corresponding to $t$ values between 0.00 and $|t|=2.14$. For larger
$|t|$ values topologically new intermediate surfaces are formed before the pinching-off of
the surface. For $|t|$ values between 2.15 and 3.32 the surface changes into a surface with
genus 5, the same value as for the G surface.
For $|t|$ values between 3.33 and 3.40 the surface changes again into a surface with
genus 29. The surface finally pinches off at $|t_1|=|t_2|=3.41$, corresponding to volume fractions
of 0.16 and 0.84.
As can be seen from Fig.\ref{fig:levsurf2}, the C(Y$^{**}$) structure exhibits
a gap for volume fractions $\phi_s$ in the range 11-38 \%,
corresponding to $t$ values between -4.00 and -1.25.
However, the largest part of the band gap map, namely for $0.17\le\phi_s\le 0.29$,
shows gaps for surfaces which have $g=5$ and not $g=29$.
The maximum bandgap width of 11.9\% is reached for $\phi_s=0.18$ ($t=-3.25$).
The corresponding midgap frequency $\omega_0=0.514$.
For this volume fraction the structure is broken apart and
consists of 20 separate units per unit cell.

For the space groups $Pm{\bar 3}m$ and $Fd{\bar 3}m$ only the level surface structures with the lowest genus,
namely the P and D structures, respectively, have a PBG. For the $I4_132$ space group,
the G structure with $g=5$ and the C(Y$^{**}$) structure with $g=29$ have a PBG.
The gap of the G structure, however, is much larger than the one of the C(Y$^{**}$) structure and is thus
far more interesting. Moreover the largest part of the band gap map of the C(Y$^{**}$) structure is not formed
by surfaces which have $g=29$ but which have $g=5$, the same genus as the gyroid.
Hence, for the space groups $Pm{\bar 3}m$, $Fd{\bar 3}m$ and
$I4_132$, the cubic structures with the simplest topology (lowest genus) have the largest PBG.
In what follows we investigate whether this is a generic property or not. We study the band gap maps of the
bicontinuous cubic structures in the order of increasing index of their cubic space groups~\cite{ITC}
(see Table V).

Two surfaces having the symmetry $Pa{\bar 3}$ but having different genus are
the $^{\pm}$Y and the C($^{\pm}$Y) surfaces.
For the $^{\pm}$Y structures with $g=21$, which exist for $0.35\le\phi_s\le 0.65$
($|t_1|=|t_2|=0.37$), no gap is found for the
range of parameters analyzed. However for the C($^{\pm}$Y) structures we find a band gap.
The variation of the band gap with volume fraction for the C($^{\pm}$Y) structures
with $g=13$ is shown in Fig.\ref{fig:levsurf2}.
The existence region of the C($^{\pm}$Y) level surface structures is $0.07\le\phi_s\le 0.93$,
corresponding to $-1.65\le t\le 1.65$. A PBG is found for volume
fractions between 0.08 and 0.37 ($-1.60\leq t\leq -0.50$) and is largest for a volume fraction of 17\%
($t=-1.30$). The largest band gap (8.1\%) is not so large as the one for the
D and G surfaces but the midgap frequency $\omega_0=0.786$ is noticeable larger.
This can be advantageous for the production of PBG crystals in the shorter
wave length regimes. Also for this space group only the dielectric structure with the lowest genus shows
a gap.

The next cubic space group we consider is the $Ia{\bar 3}$ group. The C(S) surface belongs to this
group. For the nodal approximation of the C(S) surface the morphological
image analysis method~\cite{Minsurf5} yields $g=65$. This value for the genus is much higher than $g=17$ reported
by Fisher {\sl et al.} for the minimal C(S) surface~\cite{Fisher}.
For the C(S) level surface family connected level surfaces exist for $-1.41\le t\le 1.41$, corresponding
to $\phi_s$ values between 0.23 and 0.77.
Our calculations show that the
C(S) structure has no PBG, which may be not so surprising due to the high genus and hence complex
topology of the structure.

Two other level surfaces belonging to a cubic symmetry group based on a
simple cubic Bravais lattice are the Y and C(Y) surfaces with $P4_332$ symmetry. Both level surfaces
have $g=13$. Fisher {\sl et al.} found $g=17$ for the Y surface~\cite{Fisher}.
This difference might again be caused by the fact that Fisher {\sl et al.} have worked with minimal
surfaces, while we use the nodal approximations.
The Y level surface structure, which exists for volume fractions between
0.27 and 0.73 ($-0.44\leq t\leq 0.44$), has no PBG.
For the C(Y) surface, a family of connected level surfaces exists
for $-1.84\leq t\leq 1.84$, corresponding to volume fractions in the range 8-92\%.
The band gap map for the C(Y) structure
is shown in Fig.\ref{fig:levsurf2}. The C(Y) structure exhibits a gap
for volume fractions ranging from 0.16 to 0.35 ($-1.50\leq t\leq -0.70$).
The maximum value $\Delta\omega /\omega_0 =6.2\%$ ($\omega_0 =0.630$) is reached
for a volume fraction of approximately 22\% ($t=-1.25$), as can be seen from Table V.
The Y and C(Y) level surface structures have the same symmetry, the same genus and the same dielectric contrast
but only the C(Y) structures have a PBG. The Y structures, however, only exist for a relatively high
fraction of the volume occupied by the high dielectric material,
which in general is a disadvantage for the appearance of PBGs.

A single level surface belonging to a cubic space group based on a body-centered cubic
Bravais lattice is the S surface. The S surface has $I{\bar 4}3d$ symmetry and has genus 21.
The family of connected S level surfaces exists for $-0.74\leq t\leq 0.74$, corresponding to $\phi_s$
values between 0.12 and 0.88.
Our calculations show that the S structure has no PBG for the parameters studied.

The band gap maps of the F-RD and the inverse F-RD structures
are depicted in Fig.\ref{fig:levsurf3}.
The F-RD surface has $Fm{\bar 3}m$ symmetry and genus 21. The cubic space
group $Fm{\bar 3}m$ is a supergroup of the group $Pm{\bar 3}m$,
the symmetry group of the P surface~\cite{Schnering,Fisher}. The groups
differ only by a face-centering translation.
The F-RD surfaces pinch off at $t_1=-1.12$ and $t_2=0.99$ which corresponds to values of
$\phi_s$ of 0.21 and 0.67.
The results show that the F-RD structure exhibits a gap for volume
fractions ranging from 0.25 to 0.43 ($-0.90\leq t\leq 0.00$). The maximum $\Delta\omega /\omega_0=6.6\%$
is reached approximately for a volume fraction of 31\% ($t=-0.60$). The midgap frequency
$\omega_0=0.773$.
The F-RD structure has thus a gap of comparable width as the PBG in the P structure,
but the midgap frequency is much higher.
For the inverse F-RD structure, which exists for $0.31\le\phi_s\le 0.78$
($-1.07\le t\le 1.10$), we also find a gap with an even
higher midgap frequency, namely $\omega_0=0.857$.
The maximum width of the bandgap is however smaller ($\Delta\omega /\omega_0=4.2\%$) than the one of the
direct F-RD structure.
In contrast to what is mentioned in Ref.~\cite{Minsurf4},
for this example it seems that the increase in symmetry of the dielectric structure
does not destroy the PBG property but allows structures with a higher genus to have a PBG.

Finally we study three level surfaces belonging to the cubic symmetry group $Ia{\bar 3}d$: The
I$_2$-Y$^{**}$ (called double gyroid by Wohlgemuth {\sl et al.}~\cite{Level1}),
C(I$_2$-Y$^{**}$) (belonging to the G$^{\prime}$ family~\cite{Level1})
and L (Lidinoid)~\cite{Level1} surfaces with genus $g=9$, $g=25$
and $g=33$, respectively.
The group $Ia{\bar 3}d$ contains all the symmetries of the group $I4_132$, the space group of
the G surface, as well as the inversion symmetry~\cite{ITC}.
For the I$_2$-Y$^{**}$ surface a family of connected level surfaces exists
for $-5.00\leq t\leq 2.98$, corresponding to $\phi_s$ values between
0.03 and 0.91. Band gap calculations for the direct as well as the inverse structure
show no PBG, in accordance with the result
reported in Ref.~\cite{Minsurf4} for the double gyroid structure. This result could suggest
that increasing the symmetry of the dielectric structure destroys
the appearance of a PBG (see Ref.~\cite{Minsurf4}).
The C(I$_2$-Y$^{**}$) level surface family exists from a volume fraction
$\phi_s$ of 0.09 to 0.72 corresponding to $t$ values between $t_1=-2.05$ and $t_2=0.99$.
The band gap maps for the direct and inverse
C(I$_2$-Y$^{**}$) structures are shown in Fig.\ref{fig:levsurf4}.
The results show that the C(I$_2$-Y$^{**}$) structure exhibits a gap for volume
fractions from 0.10 to 0.56 ($-2.00\leq t\leq 0.25$). The maximum $\Delta\omega /\omega_0=19.6\%$
is reached approximately for a volume fraction of 25\% ($t=-1.25$). The midgap frequency
$\omega_0=0.796$. This structure has two attractive features for the production of
a real PBG crystal: A large gap and a high midgap frequency.
This example also shows that
increasing the symmetry of the dielectric structure
does not necessarily destroy the PBG property but allows structures with a higher genus to have a PBG.
This property was also seen for the direct and inverse F-RD level surface structures.
As can be seen from Fig.\ref{fig:levsurf4}, the inverse C(I$_2$-Y$^{**}$) structure
has a much smaller PBG than the direct C(I$_2$-Y$^{**}$) structure,
namely $\Delta\omega /\omega_0=4.7\%$, and midgap frequency $\omega_0=0.675$ for a volume
fraction of 39\% ($t=0.50$).
The L level surfaces pinch-off at $t_1=-0.35$ and $t_2=0.30$ which corresponds to
values of $\phi_s$ from 0.25 to 0.50. The inverse L level structures exist
for $-0.33\le t\le 0.30$, corresponding to
volume fractions $\phi_s$ ranging between 50\% and 74\%. The direct as well as the
inverse Lidinoid structures do not show a PBG, which could be due to the relatively
high volume fractions of the high dieletric material.

\section{Tubular structures: Results}
For some $t<t_1$ or $t>t_2$ the P, D and G level surfaces form simple cubic, diamond and gyroid
lattices of disconnected units~\cite{Schnering,Grosse}.
With increasing $|t|$ the minimal surfaces first shrink and look like tubes
about the skeletal graphs. Then the tubes shrink in the middle while the ends thicken,
until the surface pinches-off.
The pinching off behavior can be controlled by considering
the skeletal graphs of the minimal surfaces using level set methods~\cite{website}.
This results in less shrinking of the tubes in the middle and less thickening of the tubes at the
ends so that more cylindrical tubular structures can be constructed.
In Table VI we give the equations for the direct and inverse dielectric tubular structures together
with 3D renderings of 2x2x2 unit cubes of the structures for various values of $t$.
Note that the equations for the direct and inverse structures are the same. The structures become
different through Eq.(2) and Eq.(3) for the direct and inverse structure, repectively.
The interval of $t$ values is chosen such that the endpoints approximate the
$t$ values for pinch-off of the surface and such that the range of $t$ values for which the structures show
a PBG is covered.
As can be seen from Tables VI the length of these
intervals can be quite different for the various tubular structures.
The corresponding volume fractions $\phi_s$ are also given. Note that tubes of dielectric materials
are in fact only seen for the inverse dielectric structures.

We study the band gap maps of direct and inverse tubular periodic dielectric structures
generated from the skeletal graphs of the P, D, and G surfaces~\cite{website}.
The P structure has sixfold and the D structure has fourfold junctions, whereas the G
tubular structure consists of triple junctions. Thus the tubular structures have the same
topology as their corresponding level surface structures. Hence $g=3$, $g=5$ and $g=9$ for
the P, G and D tubular structures, respectively.

Our calculations show that the direct tubular P structure, which exists for volume
fractions $\phi_s$ between 0.29 and 0.98 ($-19.50\leq t\leq 0.09$), has a band gap map that is similar
to the one of the P level surface structure.
As can be seen from Fig.\ref{fig:tubesP} the tubular P structure exhibits a gap
for volume fractions $\phi_s$ in the range 19-36\%, corresponding to $t$ values between -25.00
and -16.50. Note that for $-25.00<t<-19.50$ the structure is disconnected and consists of ``spheres''
on a simple cubic lattice. This also happens for the P level surface structure (see section 3).
The maximum $\Delta\omega /\omega_0=4.6\%$ is reached approximately for a volume fraction of 27\% ($t=-20.00$).
The corresponding midgap frequency $\omega_0=0.427$.
The inverse tubular P structure ($0.02\leq\phi_s\leq 0.72$, $-19.5\leq t\leq 0.09$),
which, apart from the joints
resembles the circular rod scaffold structure in the simple cubic lattice~\cite{Sozuer},
has no PBG for a dielectric contrast of 11.9. Since it has been shown that the
square and circular rod structures in the simple cubic lattice have a small PBG
for a dielectric contrast of 13~\cite{Sozuer},
our results lead to the conclusion that small changes to the structure (lower dielectric contrast and
slight shape differences) can make these small gaps disappear.

\begin{center}
\begin{figure}[t!]
\includegraphics[width=8.cm]{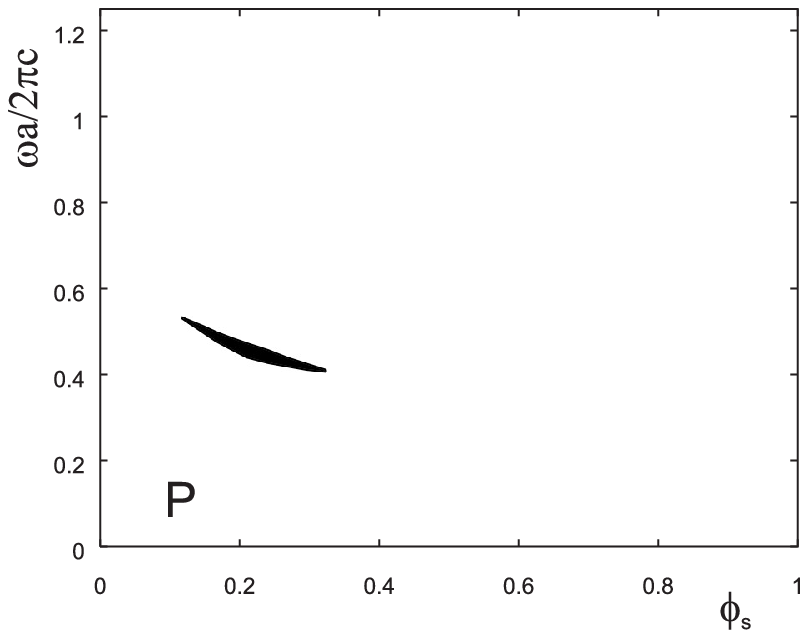}\\
\smallskip
\includegraphics[width=8.cm]{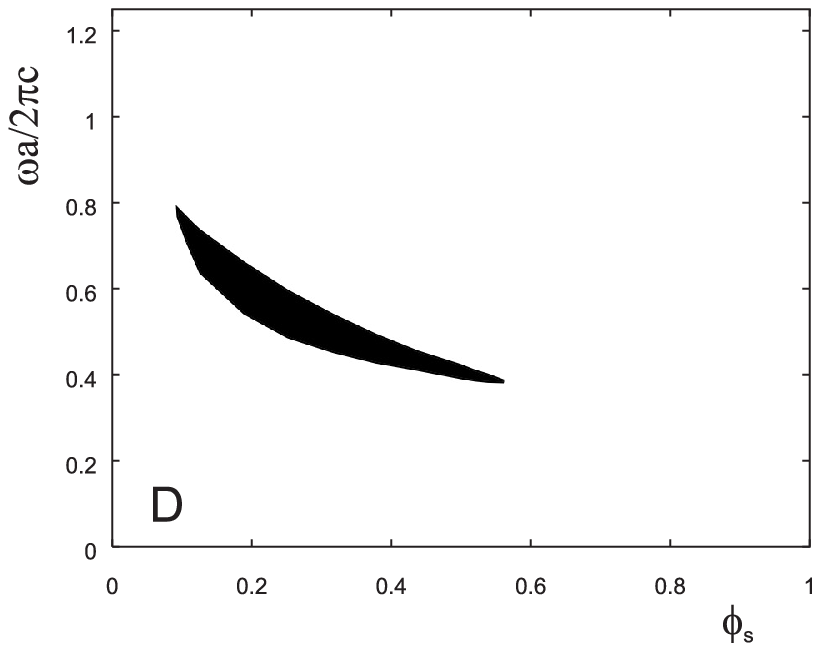}\\
\smallskip
\includegraphics[width=8.cm]{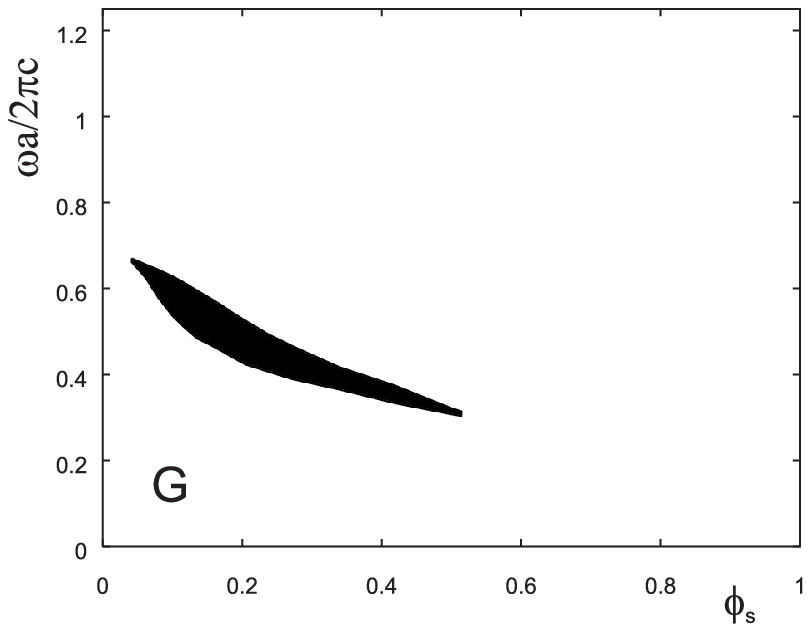}\\
\caption{Band gap maps of P, D and G level surface structures  for
$\varepsilon_s =11.9$ and $\varepsilon_b =1$.
} \label{fig:levsurf1}
\end{figure}
\end{center}

\begin{center}
\begin{figure}[t!]
\includegraphics[width=8.cm]{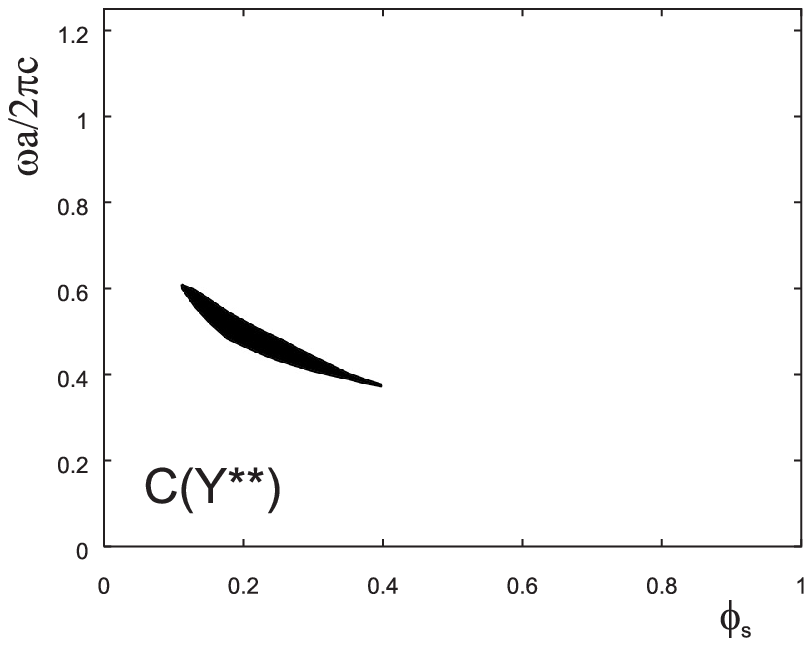}\\
\smallskip
\includegraphics[width=8.cm]{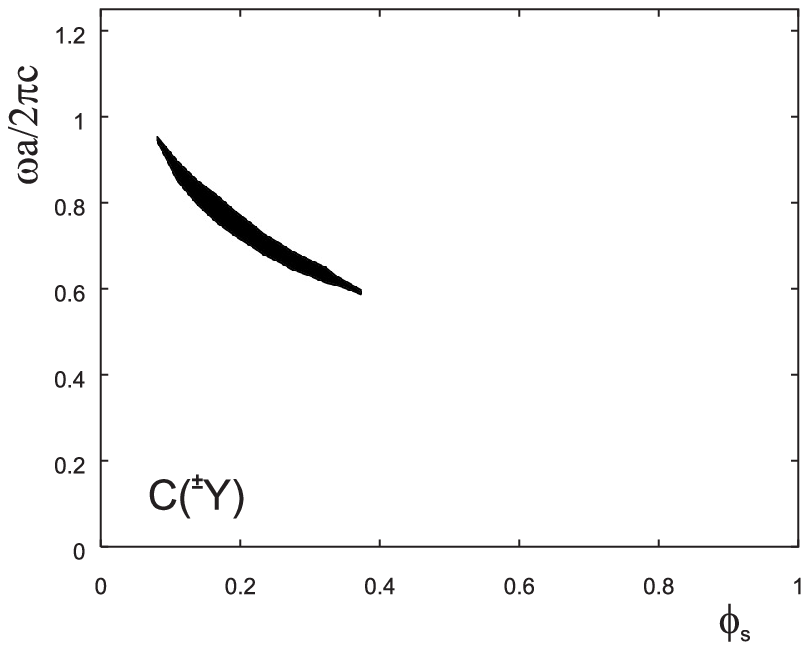}\\
\smallskip
\includegraphics[width=8.cm]{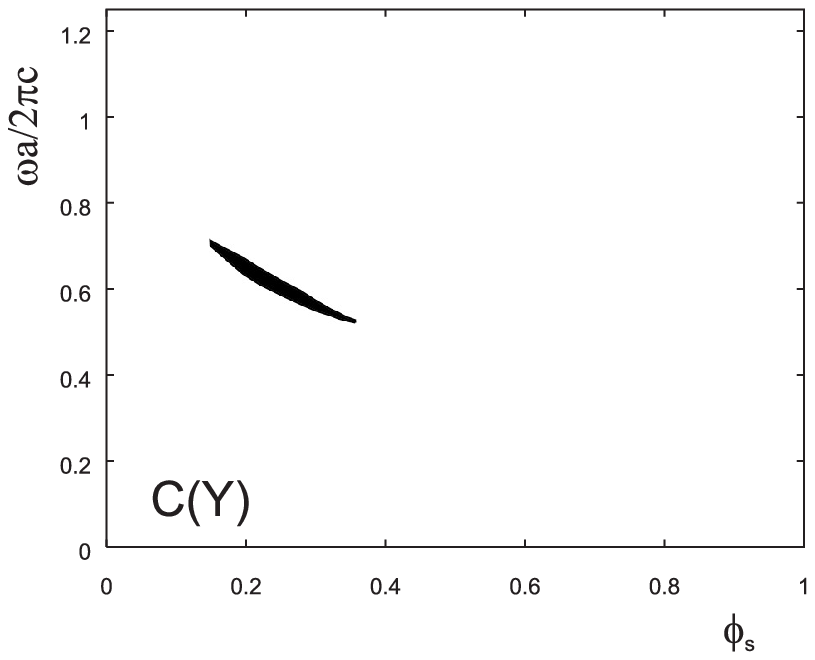}\\
\caption{Band gap maps of C(Y$^{**}$), C($^{\pm}$Y) and C(Y) level surface structures for
$\varepsilon_s =11.9$ and $\varepsilon_b =1$.
} \label{fig:levsurf2}
\end{figure}
\end{center}

\begin{center}
\begin{figure}[t!]
\includegraphics[width=8.cm]{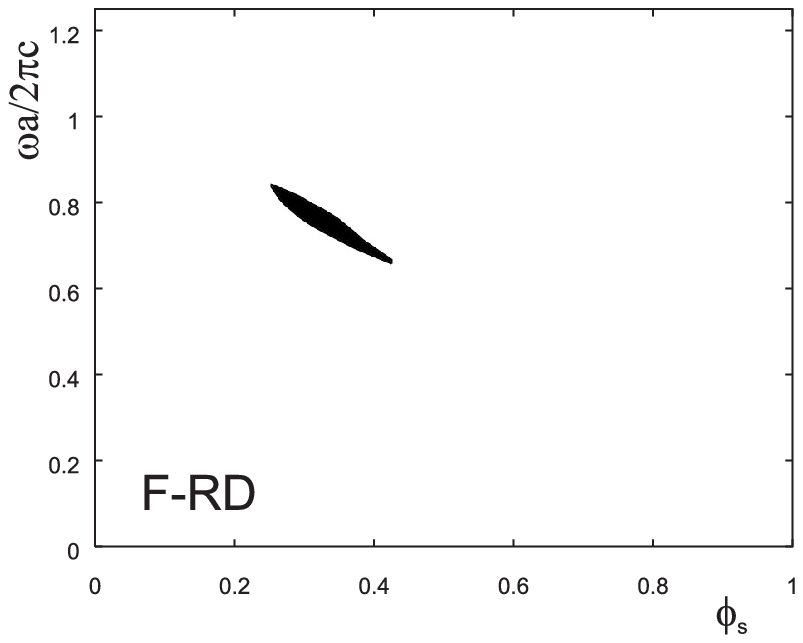}
\includegraphics[width=8.cm]{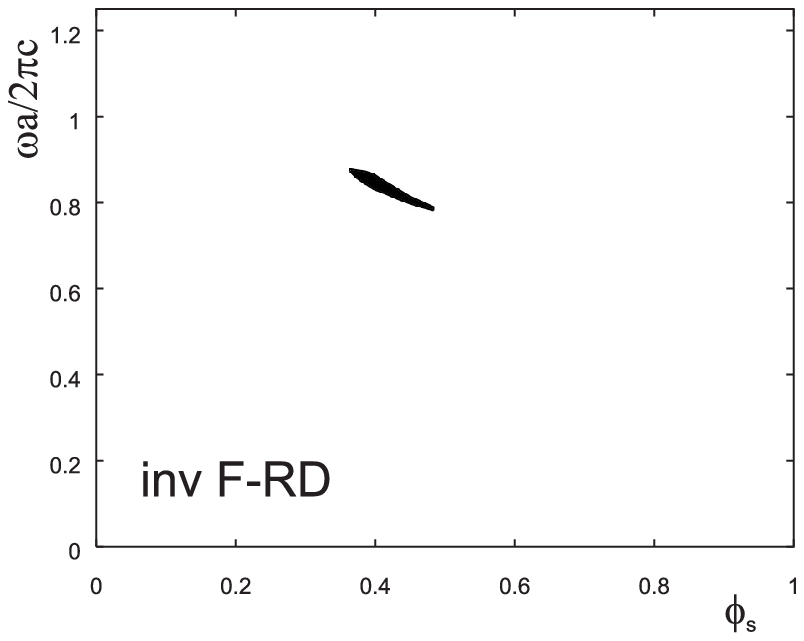}
\caption{Band gap maps of direct and inverse F-RD level surface structures for
$\varepsilon_s =11.9$ and $\varepsilon_b =1$.
} \label{fig:levsurf3}
\end{figure}
\end{center}

\begin{center}
\begin{figure}[t!]
\includegraphics[width=8.cm]{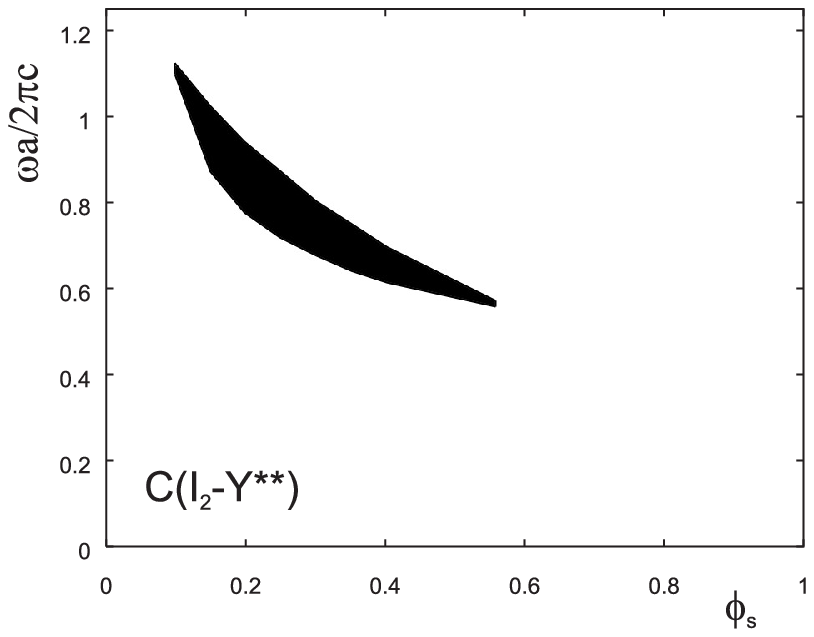}
\includegraphics[width=8.cm]{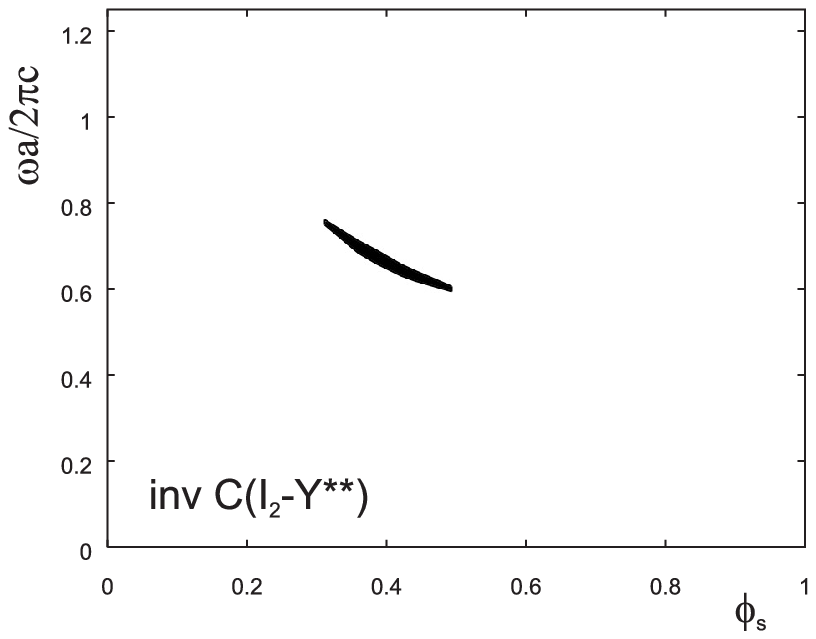}
\caption{Band gap maps of direct and inverse C(I$_2$-Y$^{**}$) level surface structures for
$\varepsilon_s =11.9$ and $\varepsilon_b =1$.
} \label{fig:levsurf4}
\end{figure}
\end{center}

\begin{center}
\begin{figure}[t!]
\includegraphics[width=8.cm]{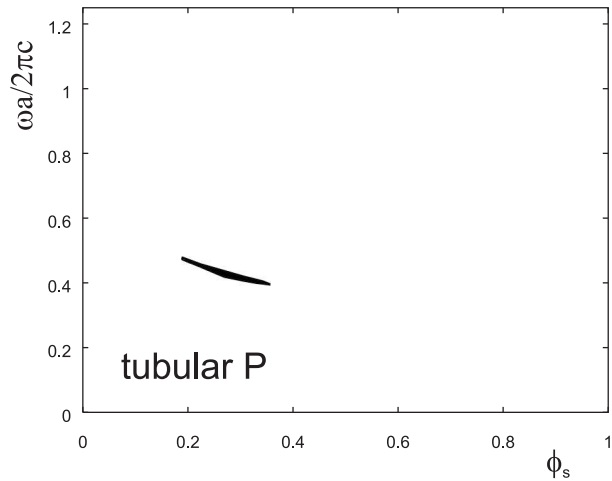}
\caption{Band gap maps of direct P tubular structures for
$\varepsilon_s =11.9$ and $\varepsilon_b =1$.
} \label{fig:tubesP}
\end{figure}
\end{center}

\begin{center}
\begin{figure}[t!]
\includegraphics[width=8.cm]{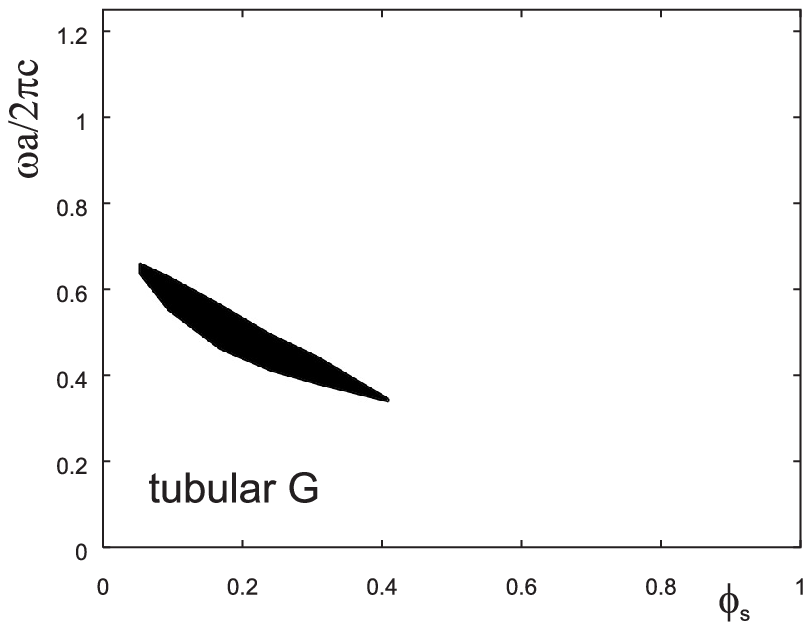}
\includegraphics[width=8.cm]{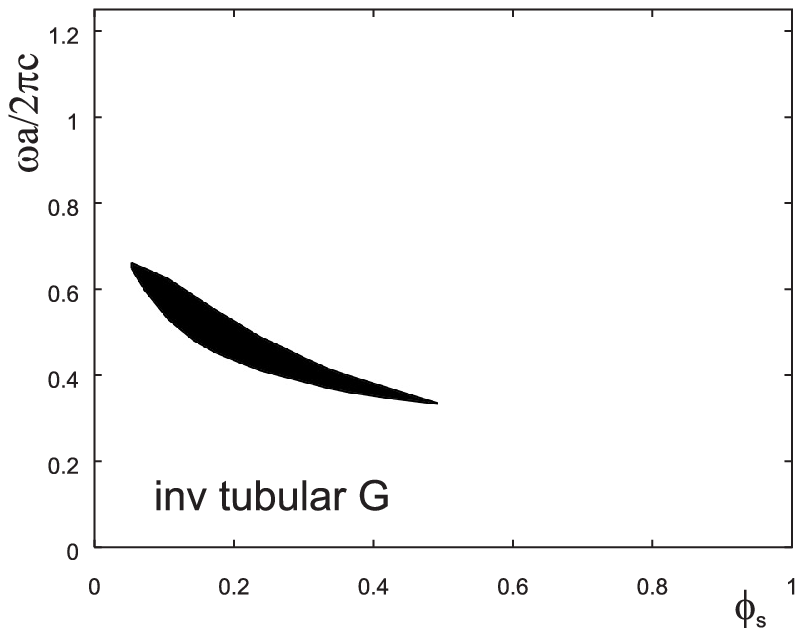}
\caption{Band gap maps of direct and inverse G tubular structures for
$\varepsilon_s =11.9$ and $\varepsilon_b =1$.
} \label{fig:tubesG}
\end{figure}
\end{center}

The direct (inverse) tubular G structure pinches off at $t_1=-27.60 (-27.60)$
and $t_2=0.59 (0.64)$, corresponding to volume fractions between 0.03 (0.01) and 0.99 (0.97).
Fig.\ref{fig:tubesG} shows the photonic band gap maps of the direct and inverse
tubular G structures. The band gap maps are similar to the band gap map of the
level surface structure (see also Fig.\ref{fig:levsurf1}).
For the direct (inverse) tubular G structure
the maximum $\Delta\omega /\omega_0=20.4\% (19.9\%)$
is reached for a volume fraction of $\phi_s=17\% (17\%)$, which corresponds to $t=-24.00 (-4.00)$.
The midgap frequency is 0.514 (0.502).

\begin{center}
\begin{figure}[t!]
\includegraphics[width=8.cm]{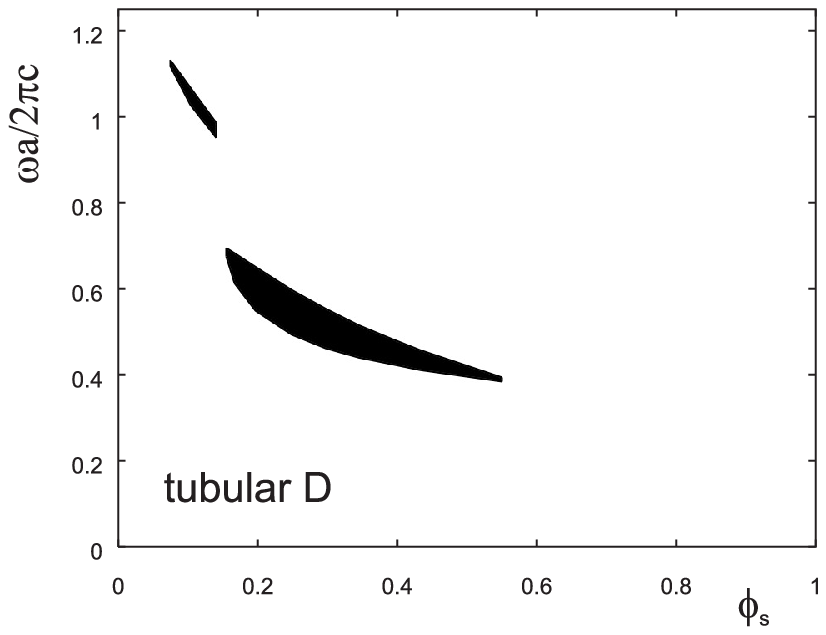}
\includegraphics[width=8.cm]{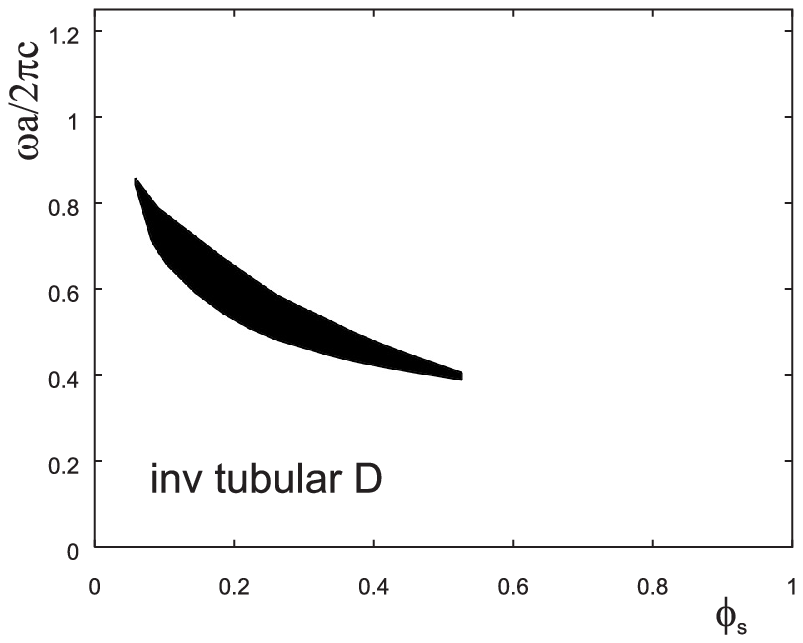}
\caption{Band gap maps of direct and inverse D tubular structures for
$\varepsilon_s =11.9$ and $\varepsilon_b =1$.
} \label{fig:tubesD}
\end{figure}
\end{center}

The band gap map of the tubular D structure is split in two, as can be seen
from Fig.\ref{fig:tubesD}. For $\phi_s<0.15$ ($t<-18.90$) the structure consists of
disconnected dielectric spherical units on a diamond lattice (see also Table VI).
This structure is known to have a PBG~\cite{Ho0}. This diamond lattice structure leads to
the upper left part of the band gap map. The right lower part of the band gap map comes from
the connected tubular structure, which exists for $-18.90\le t\le 0.88$ corresponding to
$\phi_s$ values between 0.15 and 0.88.
The midgap frequencies of the diamond lattice structure are almost twice as large as the midgap frequencies
of the connected tubular D structure, which causes the splitting in the bandgap map.
This does not occur in the case of the P level surface structure although also for this
dielectric structure the band gap map at the lowest volume fractions $\phi_s$ is formed by
disconnected ``spheres'' on a simple cubic lattice (see Fig.\ref{fig:levsurf1}).
The direct tubular D structure has a maximal band gap $\Delta\omega /\omega_0=19.2\%$
with midgap frequency $\omega_0=0.546$ for a volume fraction of
$\phi_s=25\%$, which corresponds to $t=-17.00$.
Note that the band gap map of the corresponding D level surface structure (see Fig.\ref{fig:levsurf1})
is not split in two. This is due to the fact that only the connected structures have a PBG for
certain volume fractions and that at pinch-off, the D level surface structure does not become
a diamond lattice structure.

The inverse tubular D structure has a band gap map that is similar to the band gap map
of the corresponding level surface structure, as can be seen from
Fig.\ref{fig:levsurf1} and Fig.\ref{fig:tubesD}. For the inverse D tubular structure
the maximum $\Delta\omega /\omega_0=21.4\%$
is reached for a volume fraction of $\phi_s=22\%$, which corresponds to $t=-4.00$.
The midgap frequency is 0.569.
The results are summarized in Table VII.

\section{Conclusions}
For a fixed dielectric contrast the presence
of a photonic band gap is due to a complex interplay of composition (volume fraction of structure and background),
the symmetry and topology of the periodic dielectric structure.
In general, within one symmetry group, the cubic structures
with a simple topology (low genus) show a band gap and the complex
structures with a high genus do not.
Increasing the symmetry of the dielectric structure
does not necessarily destroy the photonic band gap property but allows structures with a
higher genus to have a photonic band gap.

For the production of photonic band gap materials
the D, G, and C(I$_2$-Y$^{**}$) structures are the most interesting ones.
They all have large gaps of approximately 20\%. For the low wavelength regime
the C(I$_2$-Y$^{**}$) structure with $Ia{\bar 3}d$ symmetry might be most favourable since
its midgap frequency ($\omega_0=0.796$) is a factor of 1.5 higher than the
one for the (tubular) D and G structures.

The tubular D and G structures may be easier to produce than their level surface structure
counterparts and have similar band gap widths and midgap frequencies.

\section*{Acknowledgements}
K.M. acknowledges useful discussions with Prof. J.Th.M. De Hosson and
Prof. H. De Raedt.
This work is partially supported by the Dutch `Stichting Nationale
Computer Faciliteiten' (NCF).


\begin{table}
\begin{center}
\caption{Equations of various level surfaces~\cite{Level1,Schnering}
and 3D renderings of 2x2x2 unit cubes of level surface structures for
various values of $t$ and corresponding volume fractions $\phi_s$.
$M=2\pi m/a$ where $M=X,Y,Z$, $m=x,y,z$ and $a$ denotes
the length of the crystallographic cell.}
\label{tab1}
\begin{ruledtabular}
\begin{tabular}{ccccccccccc}
C($^{\pm}$Y)
&\multicolumn{9}{l}{
$-2\cos X\cos Y\cos Z +\sin 2X\sin  Y+\sin X\sin 2Z+\sin 2Y\sin Z=t$
}\\
\noalign{\vskip 4pt}
&\includegraphics[width=2.cm]{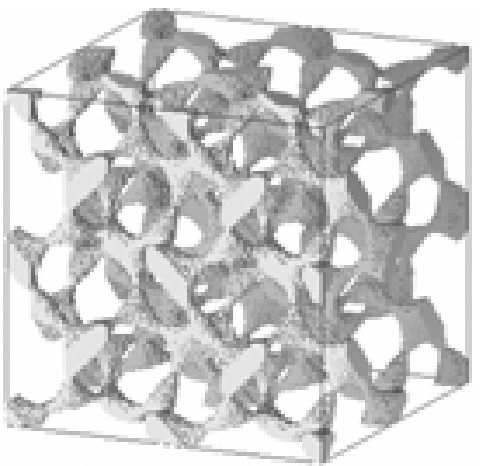}&\includegraphics[width=2.cm]{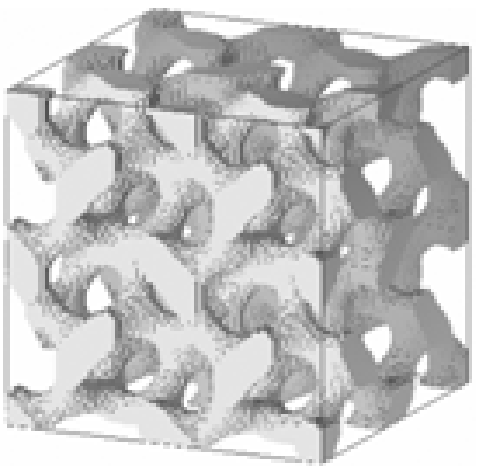}
&\includegraphics[width=2.cm]{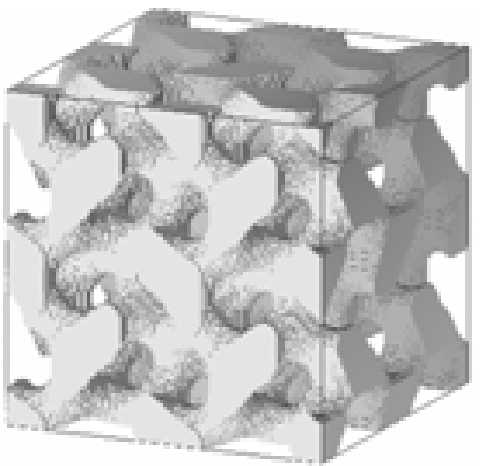}&\includegraphics[width=2.cm]{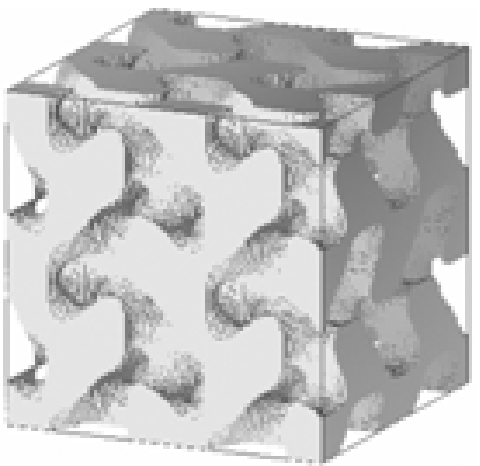}
&\includegraphics[width=2.cm]{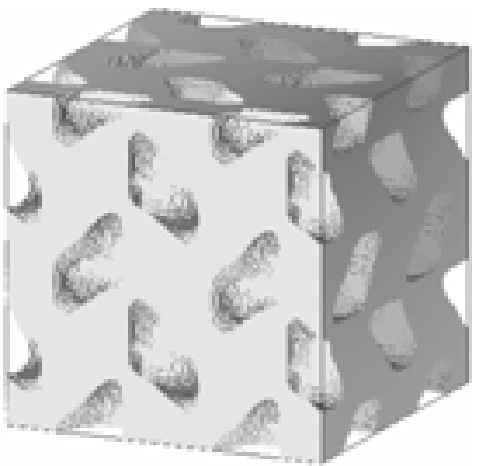}&\includegraphics[width=2.cm]{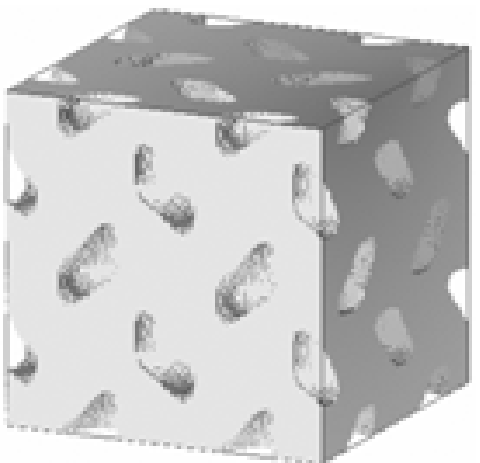}
&\includegraphics[width=2.cm]{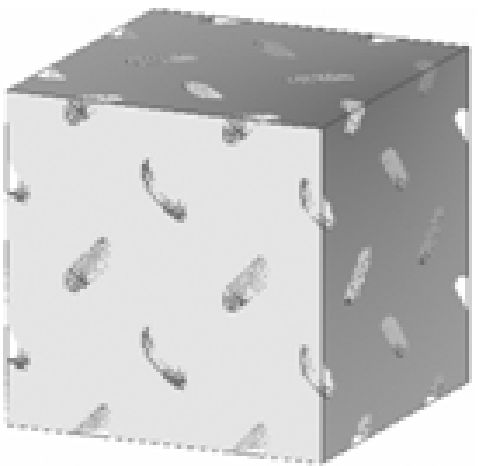}
& &\\
&$t=-1.6$&$t=-1.0$&$t=-0.5$&$t=0.0$&$t=0.5$&$t=1.0$&$t=1.6$
& &\\
&$\phi_s=0.08$&$\phi_s=0.25$&$\phi_s=0.38$&$\phi_s=0.50$&$\phi_s=0.62$&$\phi_s=0.75$&$\phi_s=0.92$
& &\\

\noalign{\vskip 2pt\hrule\vskip 4pt}
$^{\pm}Y$
&\multicolumn{9}{l}{
$2\cos X\cos Y\cos Z +\sin 2X\sin  Y+\sin X\sin 2Z+\sin 2Y\sin Z=t$
}\\ \noalign{\vskip 4pt}
&\includegraphics[width=2.cm]{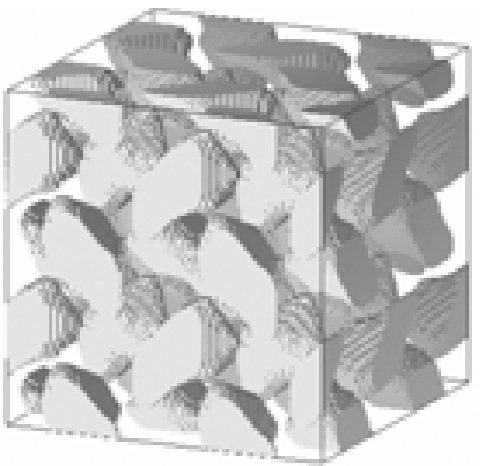}&\includegraphics[width=2.cm]{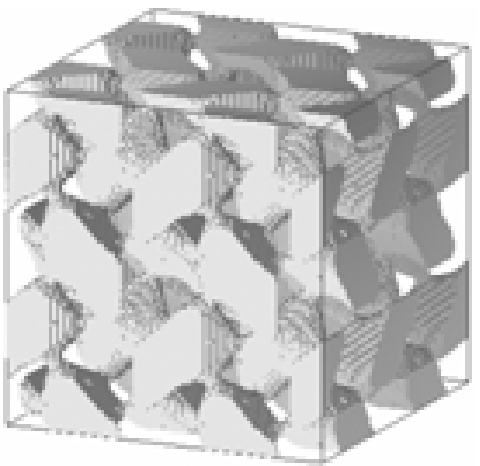}
&\includegraphics[width=2.cm]{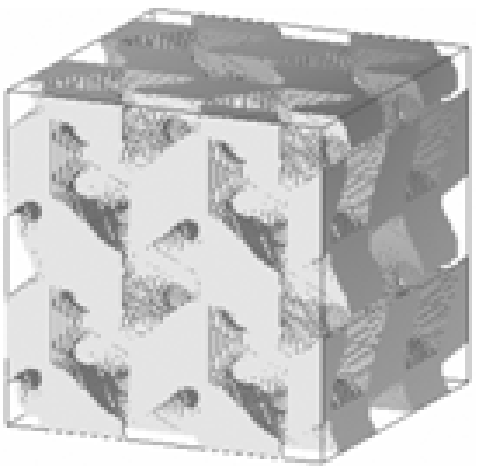}&\includegraphics[width=2.cm]{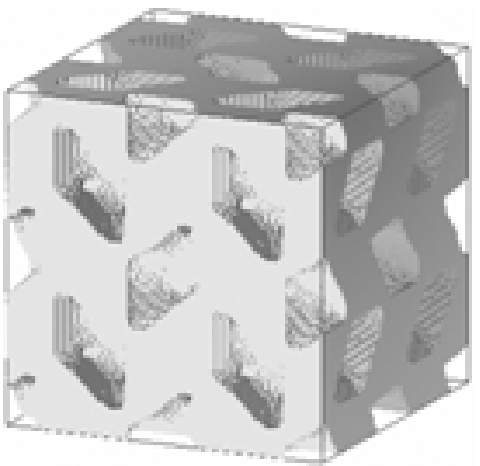}
&\includegraphics[width=2.cm]{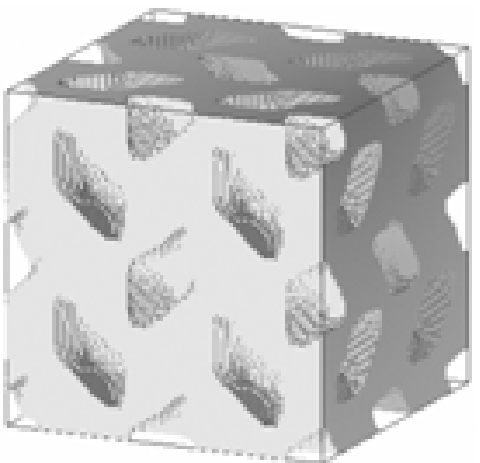}
& &\\
&$t=-0.4$&$t=-0.3$&$t=0.0$&$t=0.3$&$t=0.4$& &
& &\\ 
&$\phi_s=0.34$&$\phi_s=0.37$&$\phi_s=0.50$&$\phi_s=0.63$&$\phi_s=0.66$& &
& &\\ 
\noalign{\vskip 2pt\hrule\vskip 4pt}
C(S)
&\multicolumn{9}{l}{
$\cos 2X+\cos 2Y+\cos 2Z +2[\sin 3X\sin 2Y\cos Z+\cos X\sin 3Y\sin 2Z+\sin 2X\cos Y
\sin3Z]$}\\
&\multicolumn{9}{l}{
$+2[\sin 2X\cos 3Y\sin Z+\sin X\sin 2Y\cos 3Z +\cos 3X\sin Y\sin 2Z]=t$
}\\ \noalign{\vskip 4pt}
&\includegraphics[width=2.cm]{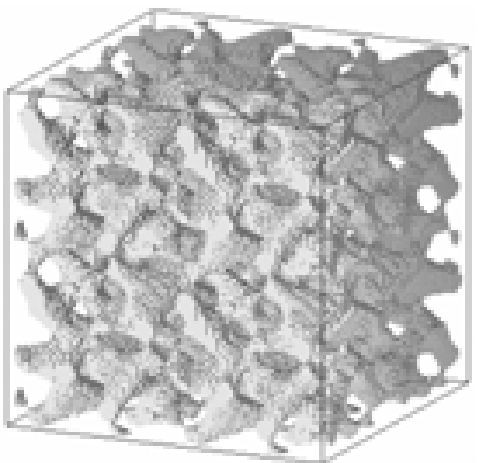}&\includegraphics[width=2.cm]{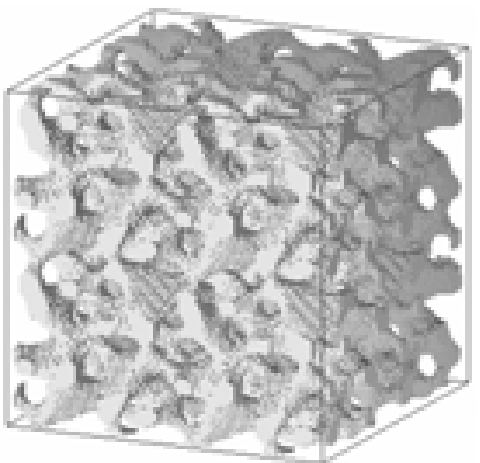}
&\includegraphics[width=2.cm]{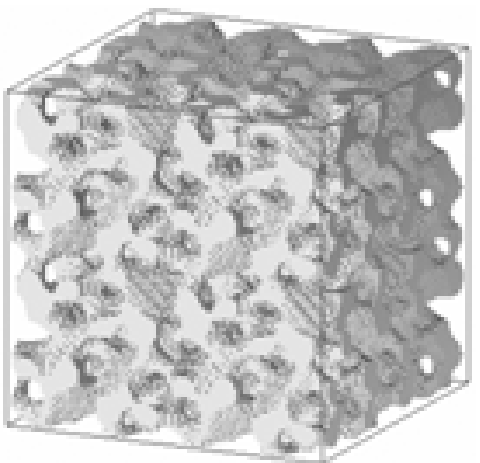}&\includegraphics[width=2.cm]{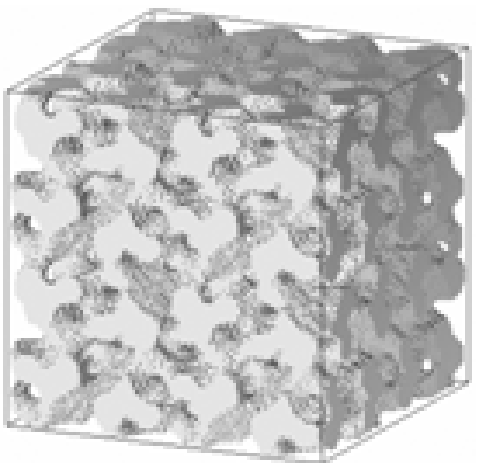}
&\includegraphics[width=2.cm]{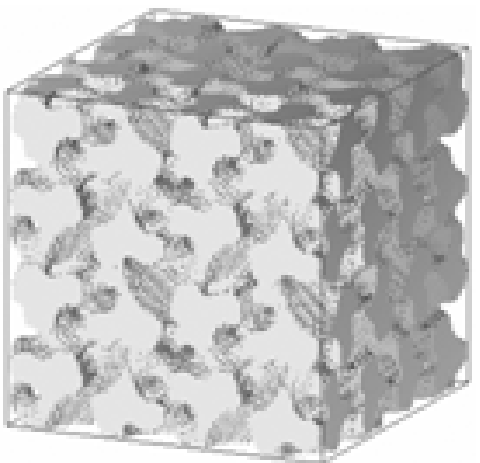}&\includegraphics[width=2.cm]{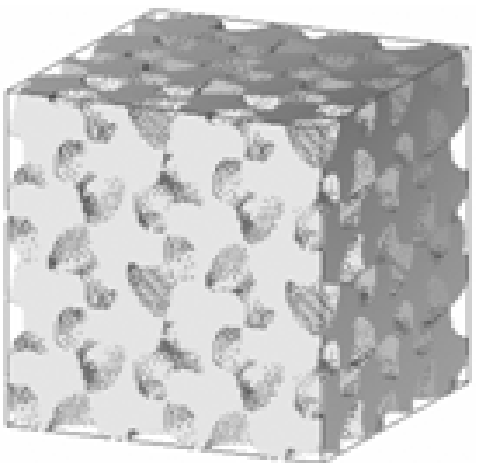}
&\includegraphics[width=2.cm]{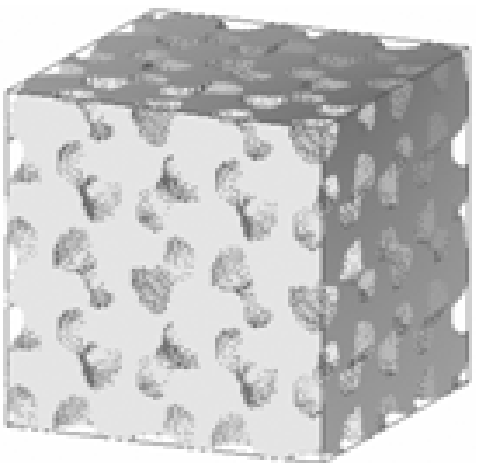}
& &\\
&$t=-1.4$&$t=-1.0$&$t=-0.5$&$t=0.0$&$t=0.5$&$t=1.0$&$t=1.4$
& &\\ 
&$\phi_s=0.24$&$\phi_s=0.30$&$\phi_s=0.40$&$\phi_s=0.50$&$\phi_s=0.60$&$\phi_s=0.70$&$\phi_s=0.76$
& &\\ 
\noalign{\vskip 2pt\hrule\vskip 4pt}
C(Y)
&\multicolumn{9}{l}{
$-\sin X\sin Y\sin Z+\sin 2X\sin Y+\sin 2Y\sin Z+\sin X\sin 2Z-\cos X\cos Y\cos Z$}\\
&\multicolumn{9}{l}{$+\sin 2X\cos Z +\cos X\sin 2Y +\cos Y\sin 2Z=t$
}\\ \noalign{\vskip 4pt}
&\includegraphics[width=2.cm]{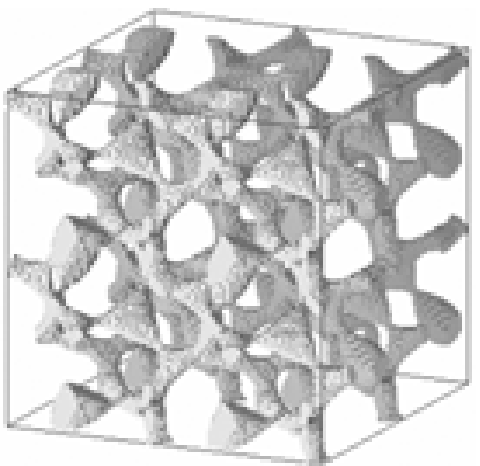}&\includegraphics[width=2.cm]{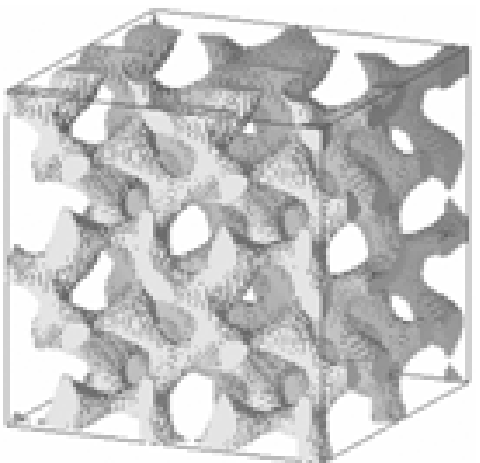}
&\includegraphics[width=2.cm]{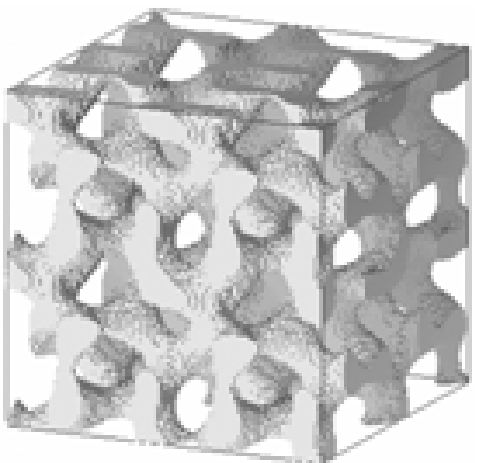}&\includegraphics[width=2.cm]{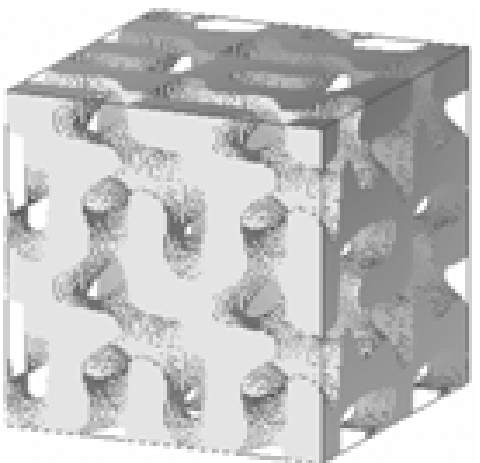}
&\includegraphics[width=2.cm]{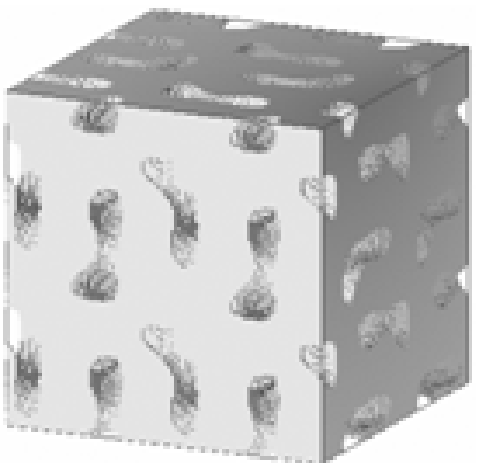}&\includegraphics[width=2.cm]{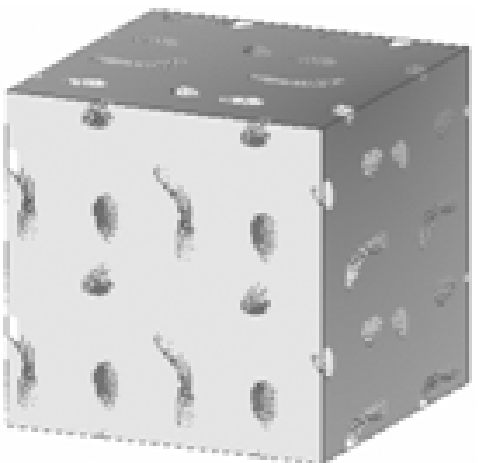}
&\includegraphics[width=2.cm]{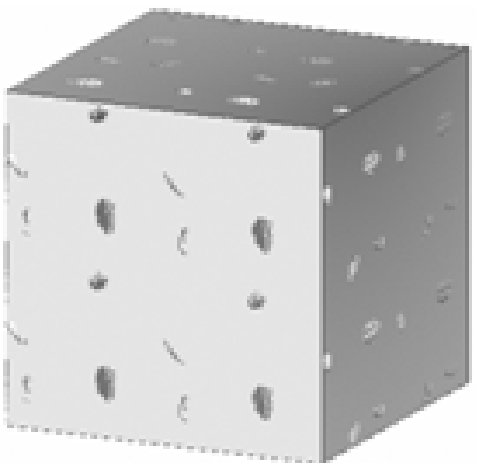}
& &\\
&$t=-1.8$&$t=-1.5$&$t=-1.0$&$t=0.0$&$t=1.0$&$t=1.5$&$t=1.8$
& &\\ 
&$\phi_s=0.09$&$\phi_s=0.16$&$\phi_s=0.28$&$\phi_s=0.50$&$\phi_s=0.72$&$\phi_s=0.84$&$\phi_s=0.91$
& &\\ 
\noalign{\vskip 2pt\hrule\vskip 4pt}
Y
&\multicolumn{9}{l}{$\cos X\cos Y\cos Z+\sin X\sin Y\sin Z+\sin 2X\sin Y+\sin 2Y\sin Z+\sin X\sin 2Z$}\\
&\multicolumn{9}{l}{$\sin 2X\cos Z +\cos X\sin 2Y +\cos Y\sin 2Z=t$
}\\ \noalign{\vskip 4pt}
&\includegraphics[width=2.cm]{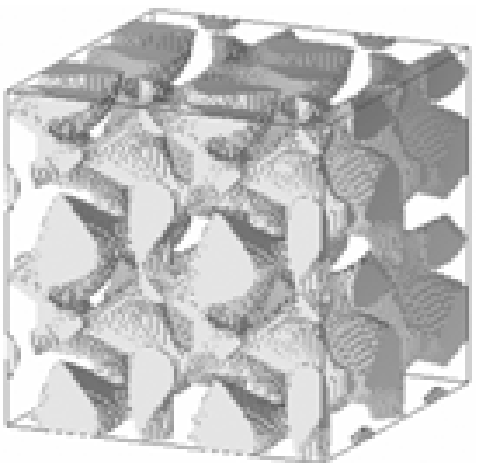}&\includegraphics[width=2.cm]{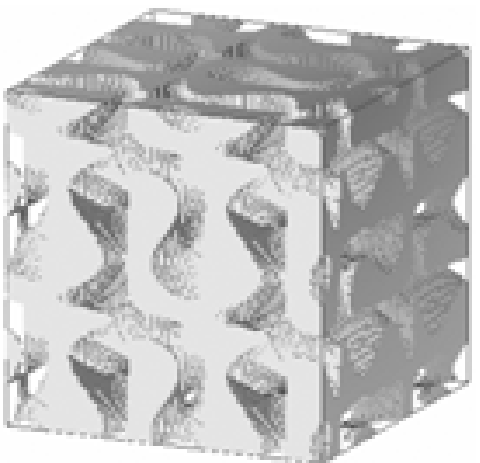}
&\includegraphics[width=2.cm]{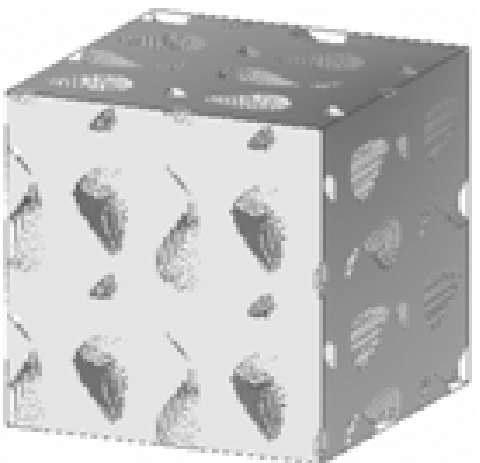}
& &\\
&$t=-0.4$&$t=0.0$&$t=0.4$& & & &
& &\\ 
&$\phi_s=0.29$&$\phi_s=0.50$&$\phi_s=0.71$& & & &
& &\\ 
\noalign{\vskip 2pt\hrule\vskip 4pt}
G
&\multicolumn{9}{l}{ 
$\sin X\cos Y+\sin Y\cos Z+\cos X\sin Z=t$
}\\ \noalign{\vskip 4pt}
&\includegraphics[width=2.cm]{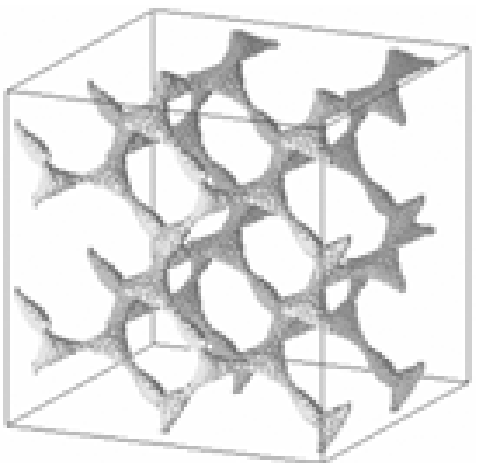}&\includegraphics[width=2.cm]{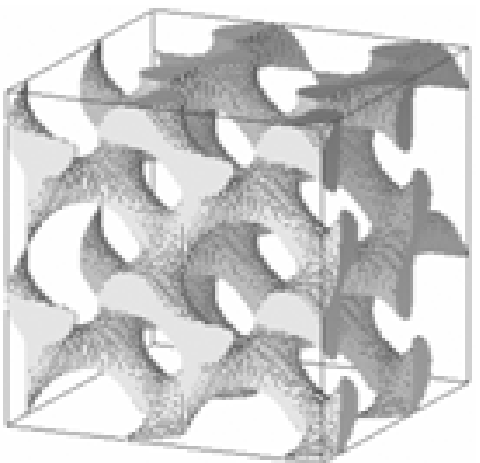}
&\includegraphics[width=2.cm]{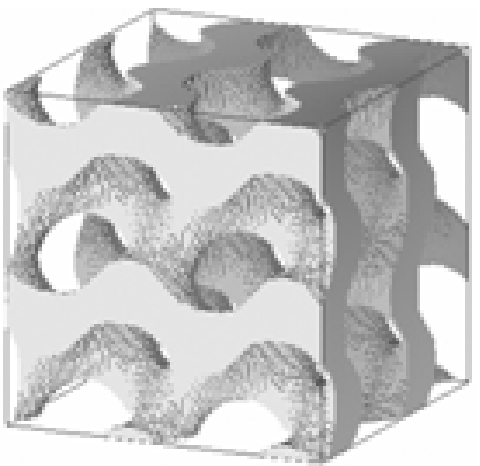}&\includegraphics[width=2.cm]{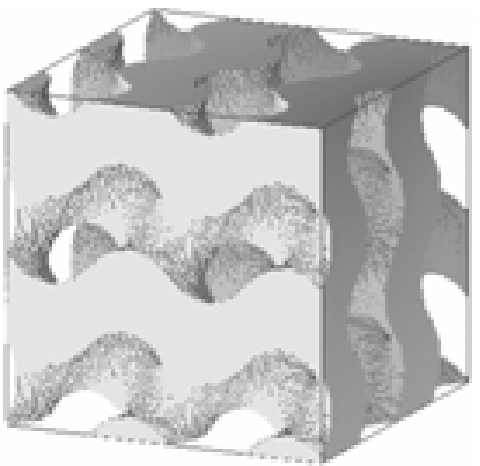}
&\includegraphics[width=2.cm]{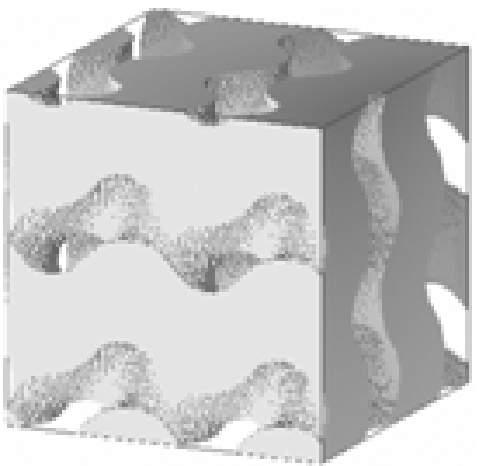}&\includegraphics[width=2.cm]{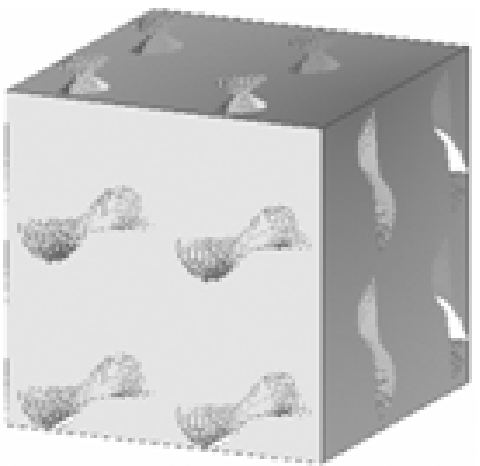}
&\includegraphics[width=2.cm]{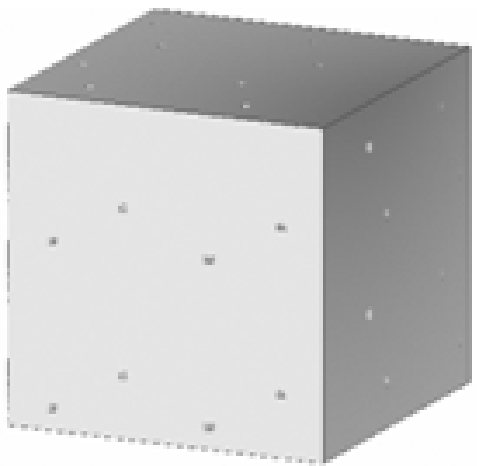}
& &\\
&$t=-1.4$&$t=-1.0$&$t=-0.5$&$t=0.0$&$t=0.5$&$t=1.0$&$t=1.4$
& &\\ 
&$\phi_s=0.02$&$\phi_s=0.17$&$\phi_s=0.34$&$\phi_s=0.50$&$\phi_s=0.66$&$\phi_s=0.83$&$\phi_s=0.98$
& &\\ 
\end{tabular}
\end{ruledtabular}
\end{center}
\end{table}
\begin{table}
\begin{center}
\caption{Same as Table 1}
\label{tab1a}
\begin{ruledtabular}
\begin{tabular}{ccccccccccc}
C(Y$^{**}$)
&\multicolumn{9}{l}{ 
$3[\sin X\cos Y+\sin Y\cos Z+\cos X\sin Z]+2[\sin 3X\cos Y+\cos X\sin 3Z +\sin 3Y\cos Z
$
}
\\
&\multicolumn{9}{l}{ 
$
-\sin X\cos 3Y-\cos 3X\sin Z-\sin Y\cos 3Z]=t$
}
\\ \noalign{\vskip 4pt}
&\includegraphics[width=2.cm]{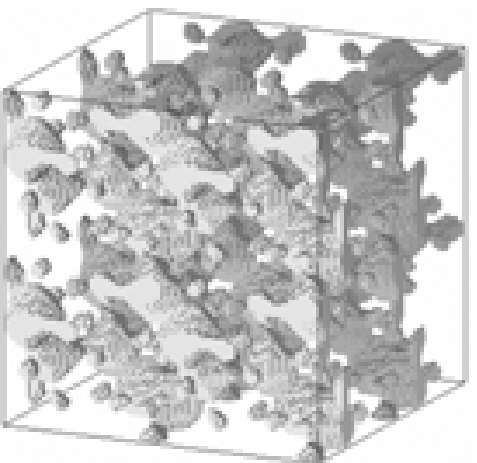}&\includegraphics[width=2.cm]{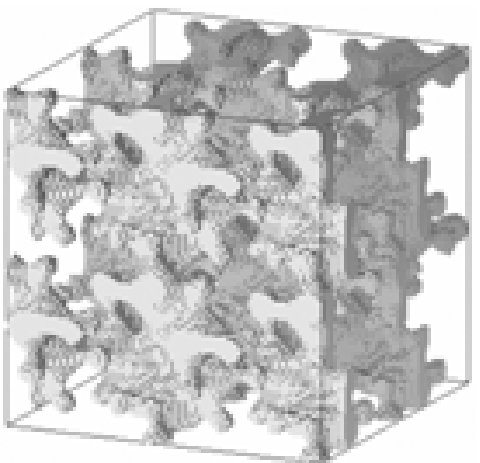}
&\includegraphics[width=2.cm]{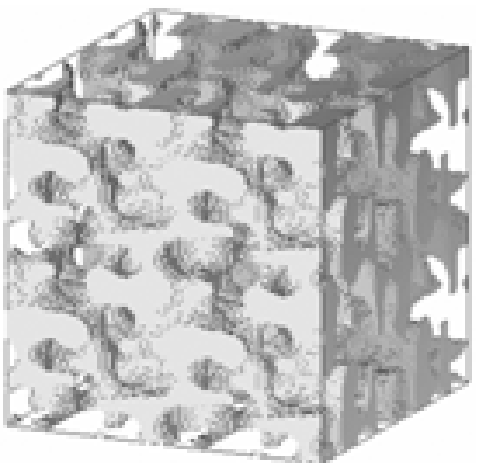}&\includegraphics[width=2.cm]{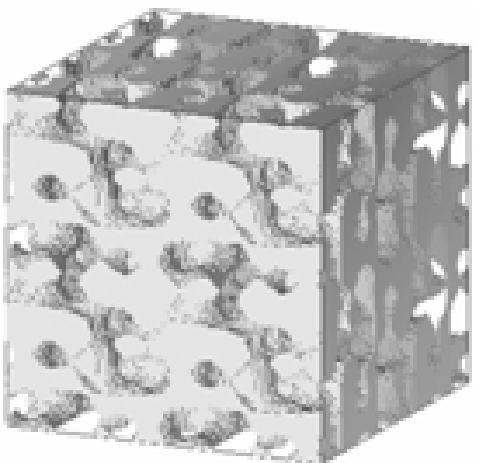}
&\includegraphics[width=2.cm]{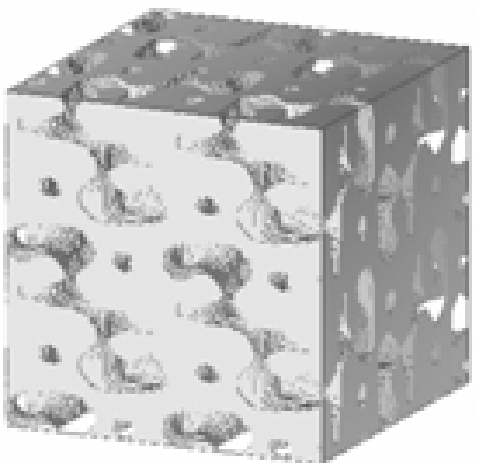}&\includegraphics[width=2.cm]{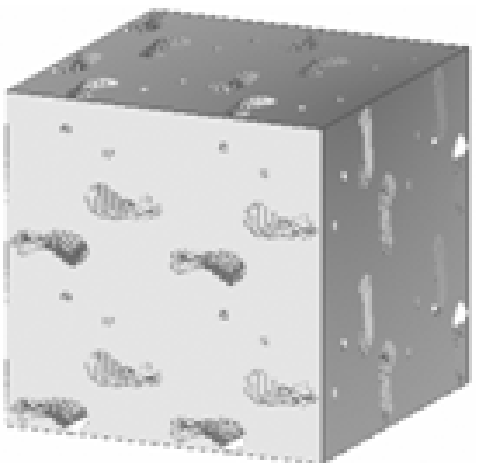}
&\includegraphics[width=2.cm]{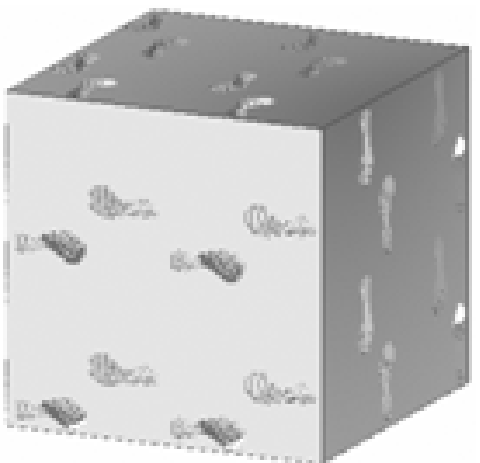}
& &\\
&$t=-3.5$&$t=-3.0$&$t=-1.0$&$t=0.0$&$t=1.0$&$t=3.0$&$t=3.5$
& &\\ 
&$\phi_s=0.15$&$\phi_s=0.20$&$\phi_s=0.40$&$\phi_s=0.50$&$\phi_s=0.60$&$\phi_s=0.80$&$\phi_s=0.85$
& &\\ 
\noalign{\vskip 2pt\hrule\vskip 4pt}
S
&\multicolumn{9}{l}{ 
$\cos 2X\sin Y\cos Z+\cos X\cos 2Y\sin Z+\sin X\cos Y\cos 2Z=t$
}\\ \noalign{\vskip 4pt}
&\includegraphics[width=2.cm]{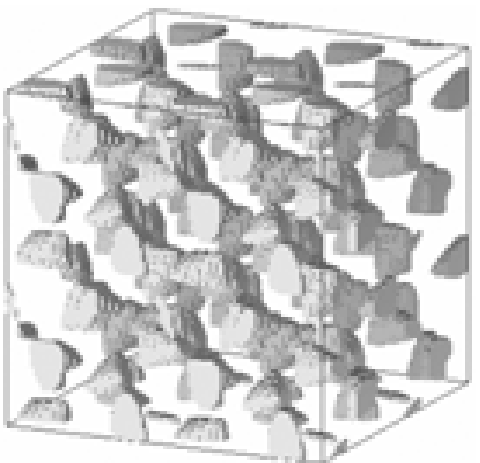}&\includegraphics[width=2.cm]{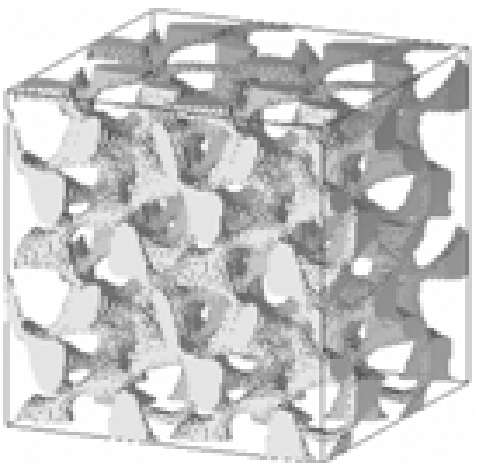}
&\includegraphics[width=2.cm]{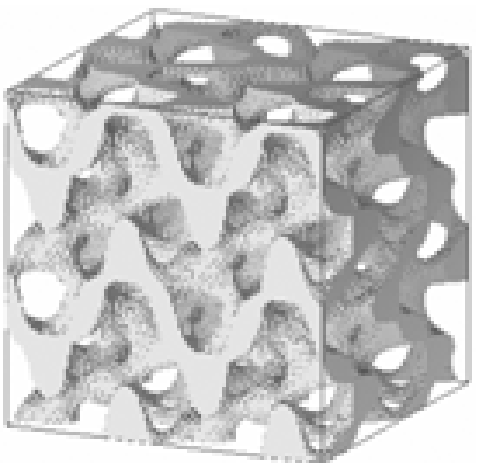}&\includegraphics[width=2.cm]{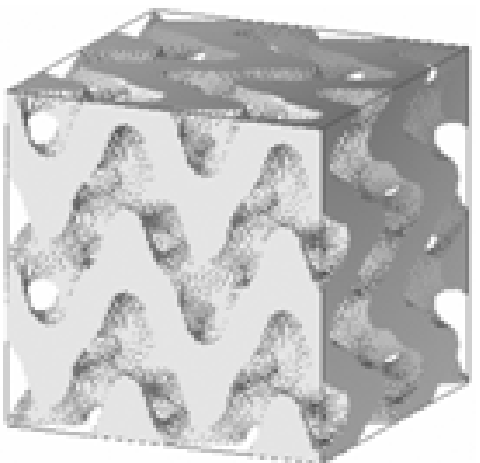}
&\includegraphics[width=2.cm]{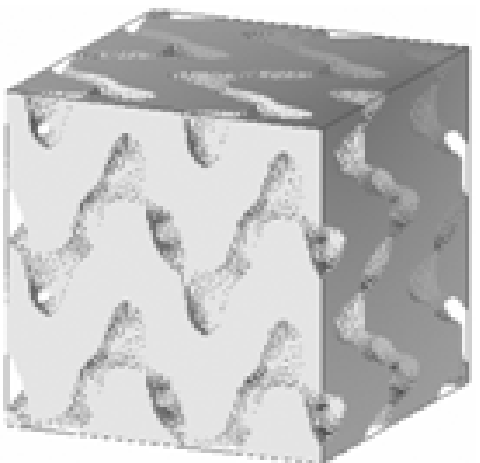}&\includegraphics[width=2.cm]{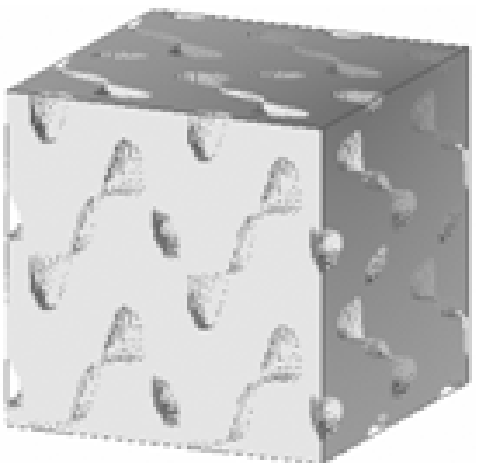}
&\includegraphics[width=2.cm]{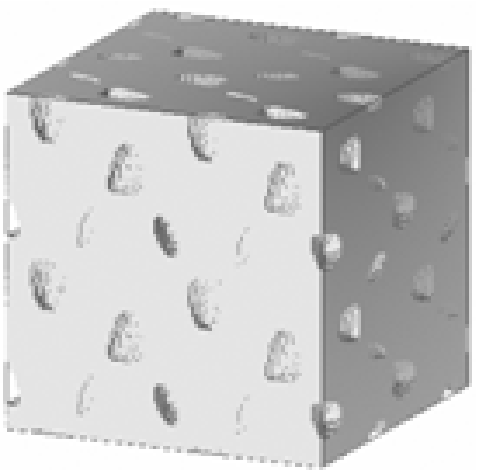}
& &\\
&$t=-0.8$&$t=-0.7$&$t=-0.5$&$t=0.0$&$t=0.5$&$t=0.7$&$t=0.8$
& &\\ 
&$\phi_s=0.09$&$\phi_s=0.14$&$\phi_s=0.25$&$\phi_s=0.50$&$\phi_s=0.75$&$\phi_s=0.86$&$\phi_s=0.91$
& &\\ 
\noalign{\vskip 2pt\hrule\vskip 4pt}
P
&\multicolumn{9}{l}{ 
$\cos X+\cos Y+\cos Z=t$
}\\ \noalign{\vskip 4pt}
&\includegraphics[width=2.cm]{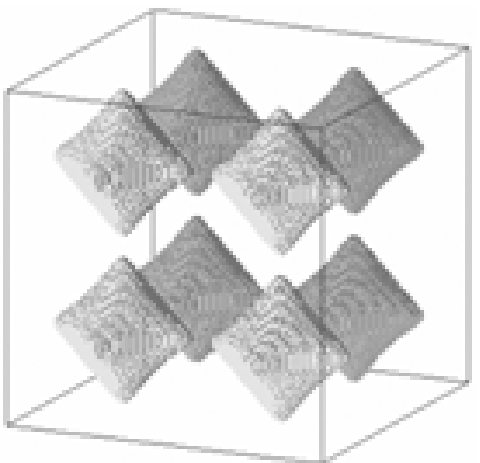}&\includegraphics[width=2.cm]{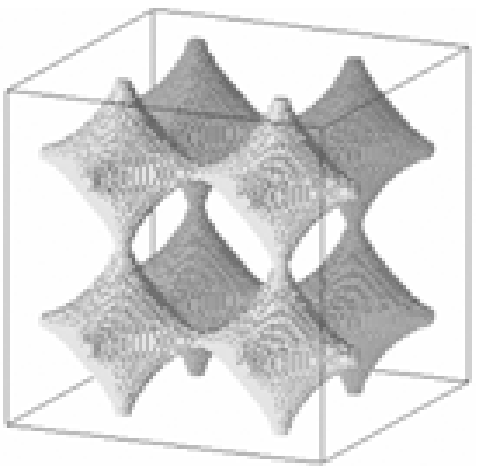}
&\includegraphics[width=2.cm]{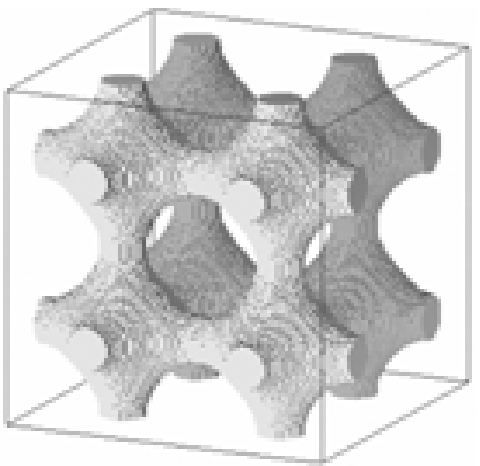}&\includegraphics[width=2.cm]{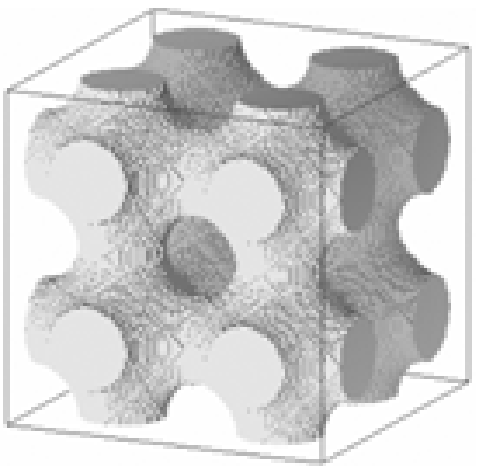}
&\includegraphics[width=2.cm]{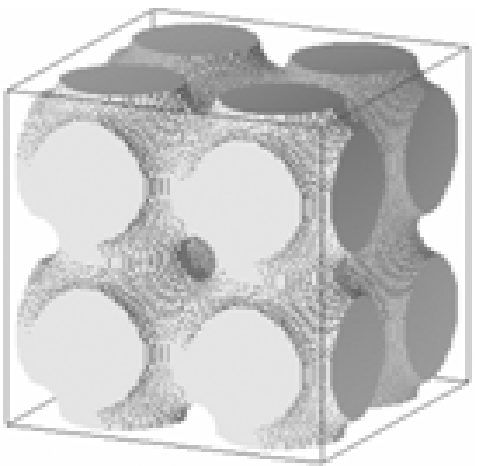}&\includegraphics[width=2.cm]{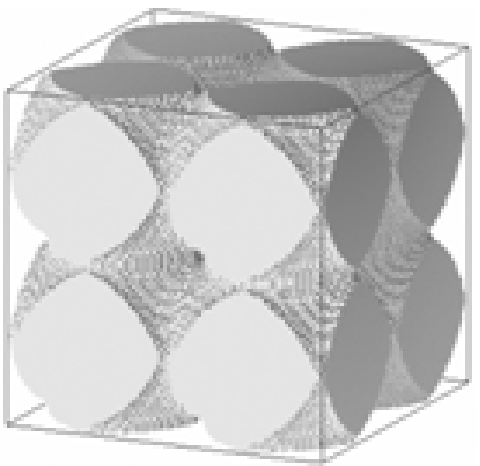}
&\includegraphics[width=2.cm]{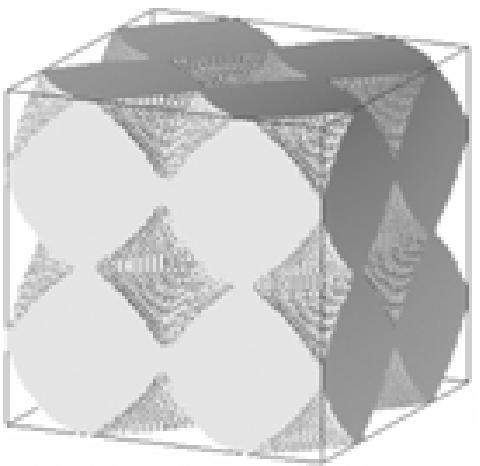}
& &\\
&$t=-1.5$&$t=-1.4$&$t=-1.0$&$t=0.0$&$t=1.0$&$t=1.4$&$t=1.5$
& &\\ 
&$\phi_s=0.13$&$\phi_s=0.16$&$\phi_s=0.26$&$\phi_s=0.50$&$\phi_s=0.74$&$\phi_s=0.84$&$\phi_s=0.87$
& &\\ 
\noalign{\vskip 2pt\hrule\vskip 4pt}
I-WP
&\multicolumn{9}{l}{ 
$2[\cos X\cos Y+\cos Y\cos Z+\cos X\cos Z]-[\cos 2X +\cos 2Y +\cos 2Z]=t$
}\\ \noalign{\vskip 4pt}
&\includegraphics[width=2.cm]{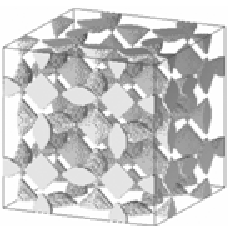}&\includegraphics[width=2.cm]{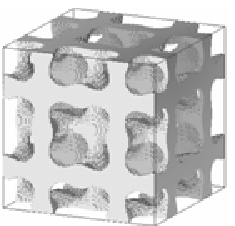}
&\includegraphics[width=2.cm]{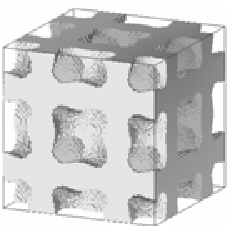}&\includegraphics[width=2.cm]{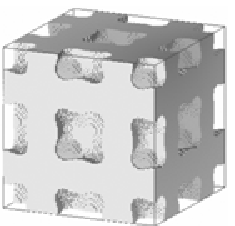}
&\includegraphics[width=2.cm]{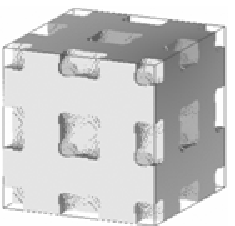}&\includegraphics[width=2.cm]{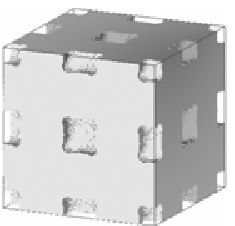}
&\includegraphics[width=2.cm]{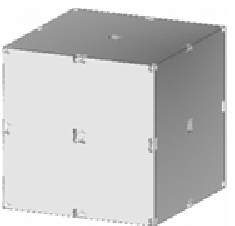}
& &\\
&$t=-3.0$&$t=-2.0$&$t=-1.0$&$t=0.0$&$t=1.0$&$t=2.0$&$t=2.98$
& &\\ 
&$\phi_s=0.10$&$\phi_s=0.22$&$\phi_s=0.34$&$\phi_s=0.47$&$\phi_s=0.61$&$\phi_s=0.76$&$\phi_s=0.99$
& &\\ 
\noalign{\vskip 2pt\hrule\vskip 4pt}
inv I-WP
&\multicolumn{9}{l}{ 
$2[\cos X\cos Y+\cos Y\cos Z+\cos X\cos Z]-[\cos 2X +\cos 2Y +\cos 2Z]=t$
}\\ \noalign{\vskip 4pt}
&\includegraphics[width=2.cm]{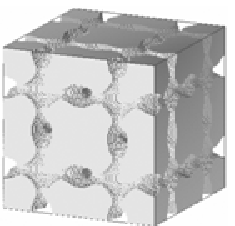}&\includegraphics[width=2.cm]{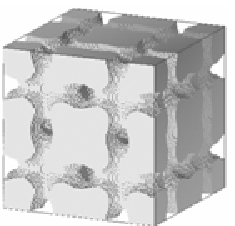}
&\includegraphics[width=2.cm]{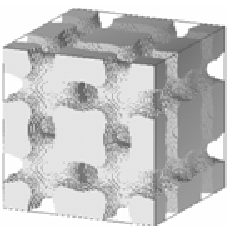}&\includegraphics[width=2.cm]{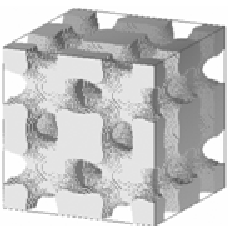}
&\includegraphics[width=2.cm]{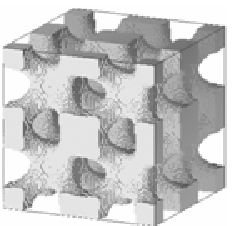}&\includegraphics[width=2.cm]{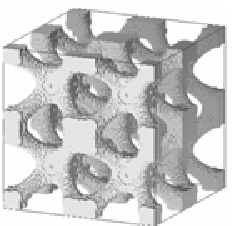}
&\includegraphics[width=2.cm]{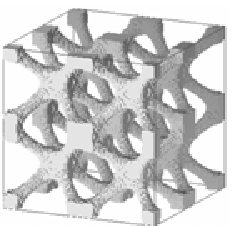}
& &\\
&$t=-2.5$&$t=-2.0$&$t=-1.0$&$t=0.0$&$t=1.0$&$t=2.0$&$t=2.5$
& &\\ 
&$\phi_s=0.84$&$\phi_s=0.78$&$\phi_s=0.66$&$\phi_s=0.53$&$\phi_s=0.39$&$\phi_s=0.24$&$\phi_s=0.14$
& &\\ 
\noalign{\vskip 2pt\hrule\vskip 4pt}
C(P)
&\multicolumn{9}{l}{ 
$\cos X+\cos Y+\cos Z +4\cos X\cos Y\cos Z=t$
}\\ \noalign{\vskip 4pt}
&\includegraphics[width=2.cm]{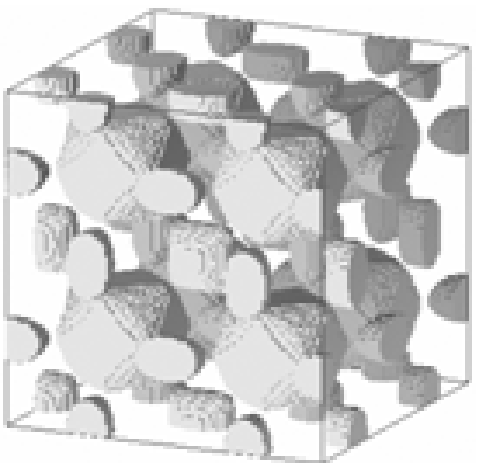}&\includegraphics[width=2.cm]{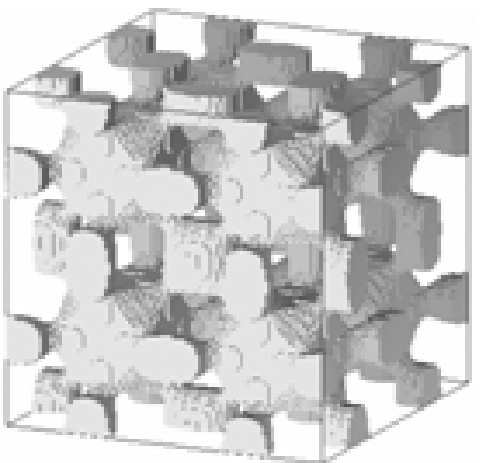}
&\includegraphics[width=2.cm]{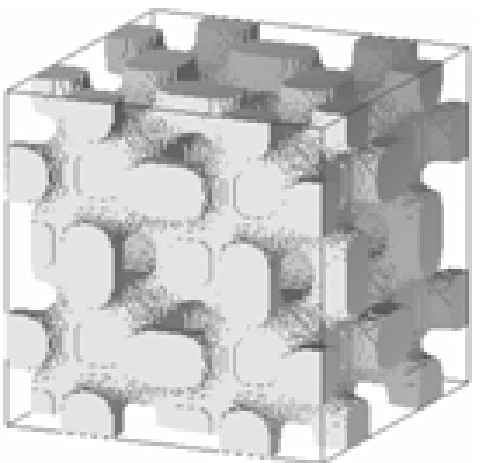}&\includegraphics[width=2.cm]{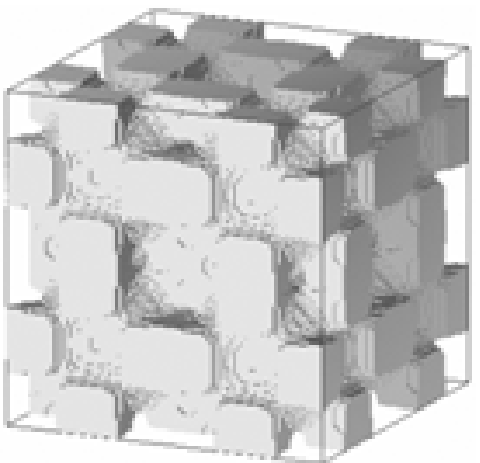}
&\includegraphics[width=2.cm]{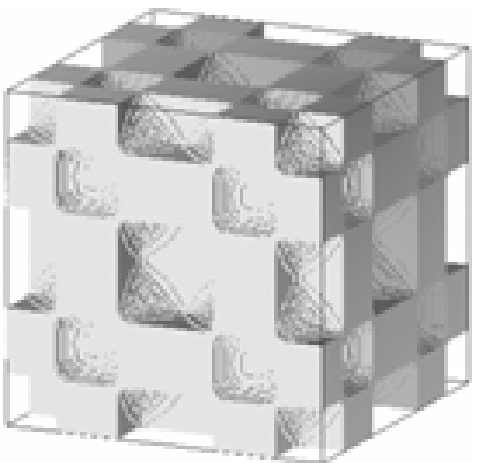}& %
& %
& &\\ 
&$t=-0.7$&$t=-0.5$&$t=0.0$&$t=0.5$&$t=0.7$& &
& &\\ 
&$\phi_s=0.27$&$\phi_s=0.33$&$\phi_s=0.50$&$\phi_s=0.67$&$\phi_s=0.73$& &
& &\\ 
\end{tabular}
\end{ruledtabular}
\end{center}
\end{table}
\begin{table}
\begin{center}
\caption{Same as Table 1}
\label{tab1b}
\begin{ruledtabular}
\begin{tabular}{ccccccccccc}
F-RD
&\multicolumn{9}{l}{ 
$4\cos X\cos Y\cos Z -[\cos 2X\cos 2Y+\cos 2X\cos 2Z+\cos 2Y\cos 2Z]=t$
}\\ \noalign{\vskip 4pt}
&\includegraphics[width=2.cm]{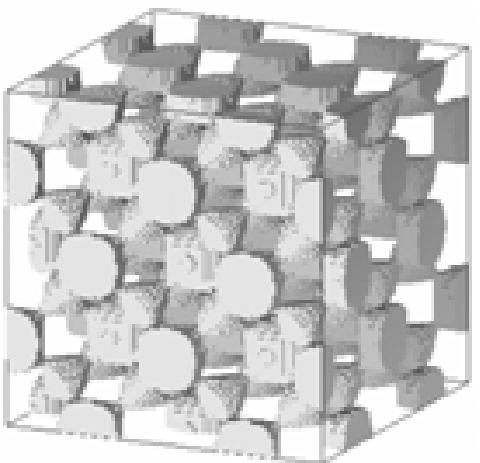}&\includegraphics[width=2.cm]{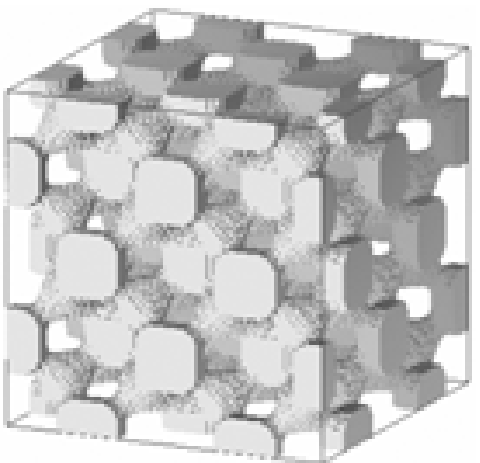}
&\includegraphics[width=2.cm]{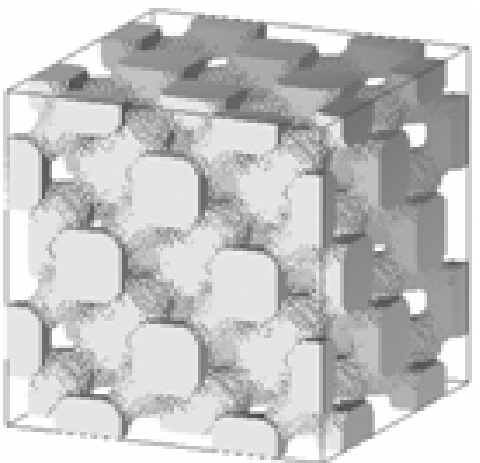}&\includegraphics[width=2.cm]{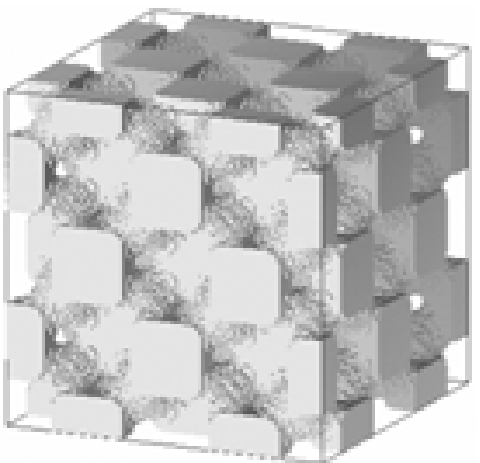}
&\includegraphics[width=2.cm]{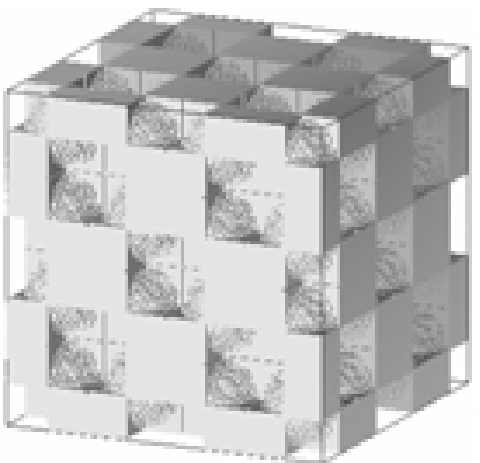}
& &\\
&$t=-1.0$&$t=-0.5$&$t=0.0$&$t=0.5$&$t=0.7$& &
& &\\ 
&$\phi_s=0.23$&$\phi_s=0.33$&$\phi_s=0.43$&$\phi_s=0.54$&$\phi_s=0.60$& &
& &\\ 
\noalign{\vskip 2pt\hrule\vskip 4pt}
inv F-RD
&\multicolumn{9}{l}{ 
$4\cos X\cos Y\cos Z -[\cos 2X\cos 2Y+\cos 2X\cos 2Z+\cos 2Y\cos 2Z]=t$
}\\ \noalign{\vskip 4pt}
&\includegraphics[width=2.cm]{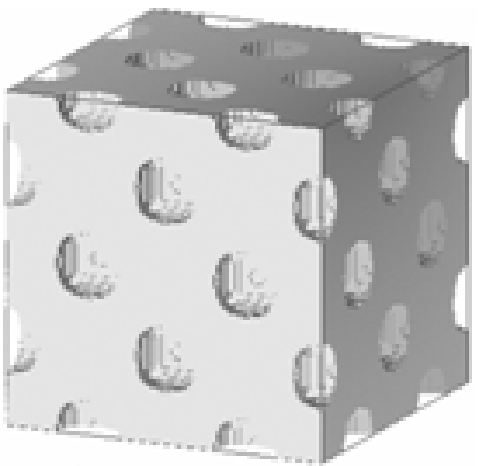}&\includegraphics[width=2.cm]{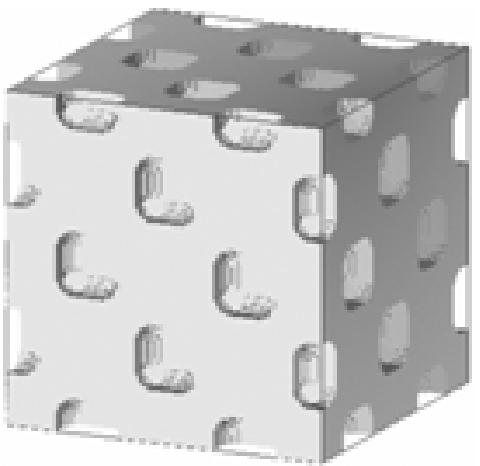}
&\includegraphics[width=2.cm]{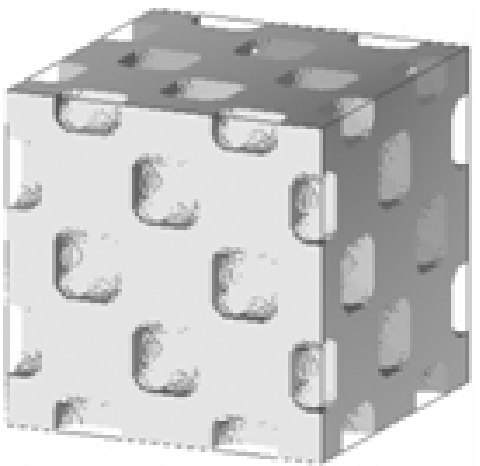}&\includegraphics[width=2.cm]{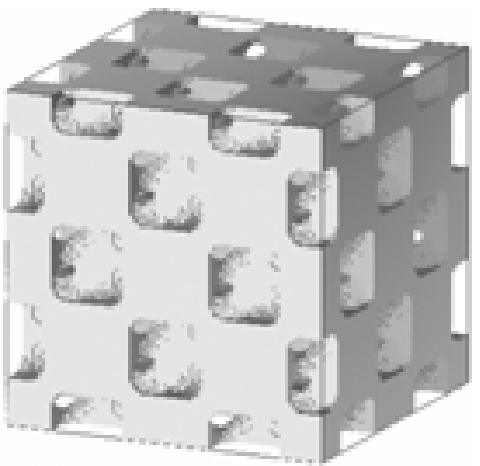}
&\includegraphics[width=2.cm]{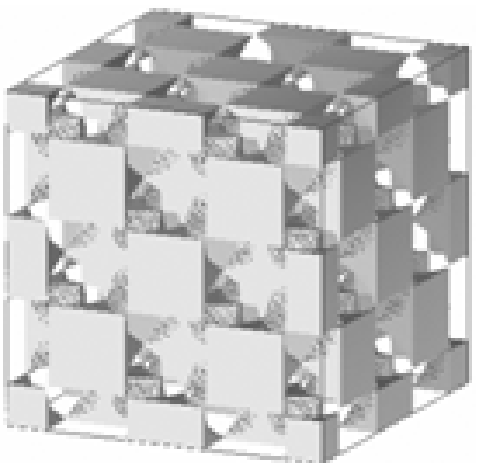}
& &\\
&$t=-1.0$&$t=-0.5$&$t=0.0$&$t=0.5$&$t=1.0$& &
& &\\ 
&$\phi_s=0.77$&$\phi_s=0.67$&$\phi_s=0.57$&$\phi_s=0.46$&$\phi_s=0.33$& &
& &\\ 
\noalign{\vskip 2pt\hrule\vskip 4pt}
D
&\multicolumn{9}{l}{ 
$\sin X\sin Y\sin Z +\sin X\cos Y\cos Z +\cos X\sin Y\cos Z +\cos X\cos Y\sin Z=t$
}\\ \noalign{\vskip 4pt}
&\includegraphics[width=2.cm]{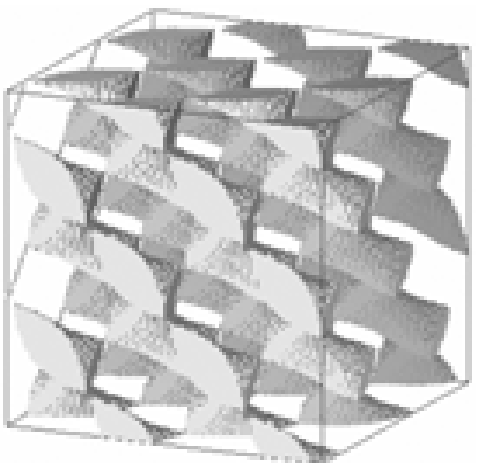}&\includegraphics[width=2.cm]{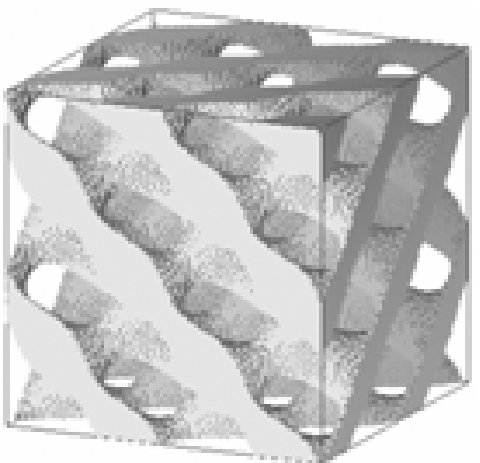}
&\includegraphics[width=2.cm]{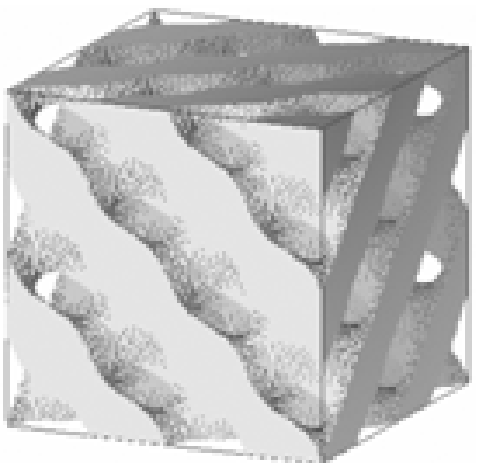}&\includegraphics[width=2.cm]{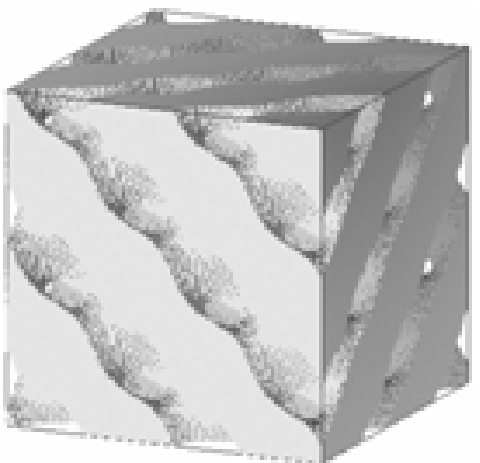}
&\includegraphics[width=2.cm]{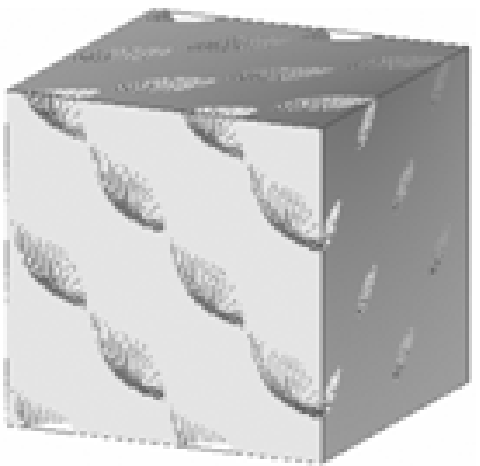}
& &\\
&$t=-1.0$&$t=-0.5$&$t=0.0$&$t=0.5$&$t=1.0$& &
& &\\ 
&$\phi_s=0.16$&$\phi_s=0.33$&$\phi_s=0.50$&$\phi_s=0.67$&$\phi_s=0.84$& &
& &\\ 
\noalign{\vskip 2pt\hrule\vskip 4pt}
C(D)
&\multicolumn{9}{l}{ 
$\cos (3X+Y)\cos Z -\sin (3X-Y)\sin Z +\cos (X+3Y)\cos Z +\sin (X-3Y)\sin Z $}\\
&\multicolumn{9}{l}{
$ +\cos (X-Y)\cos 3Z -\sin (X+Y)\sin 3Z=t$
}\\ \noalign{\vskip 4pt}
&\includegraphics[width=2.cm]{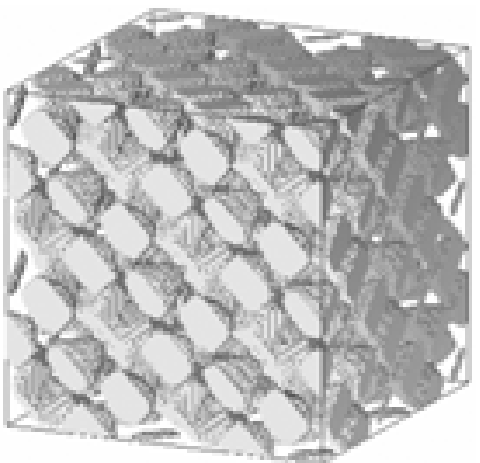}&\includegraphics[width=2.cm]{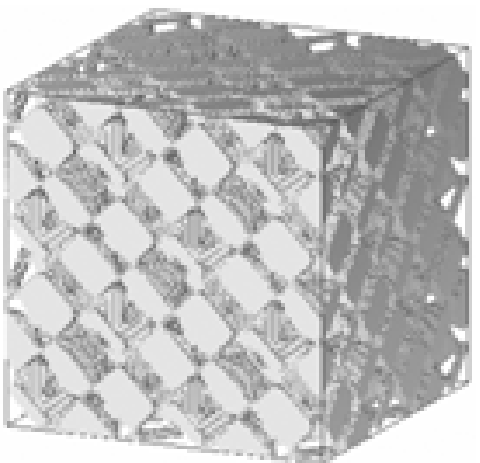}
&\includegraphics[width=2.cm]{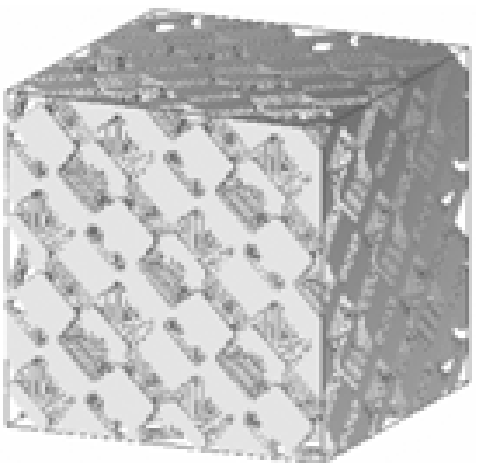}&\includegraphics[width=2.cm]{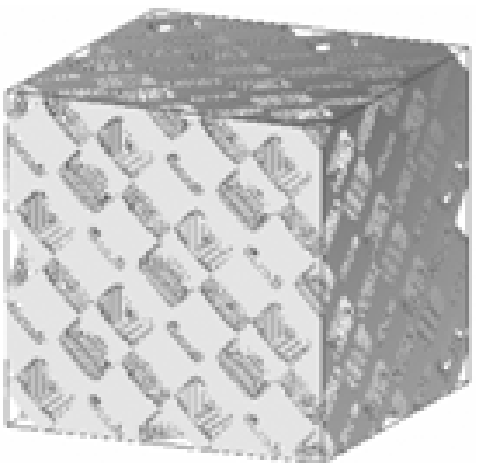}
&\includegraphics[width=2.cm]{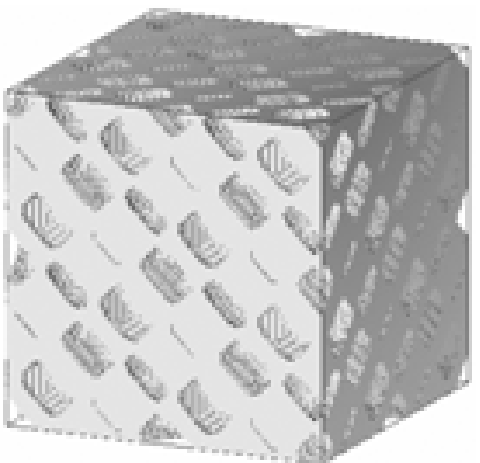}
& &\\
&$t=-0.3$&$t=-0.1$&$t=0.0$&$t=0.1$&$t=0.3$& &
& &\\ 
&$\phi_s=0.34$&$\phi_s=0.40$&$\phi_s=0.46$&$\phi_s=0.56$&$\phi_s=0.59$& &
& &\\ 
\noalign{\vskip 2pt\hrule\vskip 4pt}
I$_2$ - Y$^{**}$
&\multicolumn{9}{l}{ 
$-2[\sin 2X\cos Y\sin Z+\sin X\sin 2Y\cos Z+\cos X \sin Y\sin 2Z]$}\\
&\multicolumn{9}{l}{
$+\cos 2X\cos 2Y+\cos 2Y\cos 2Z+\cos 2X\cos 2Z=t$
}\\ \noalign{\vskip 4pt}
&\includegraphics[width=2.cm]{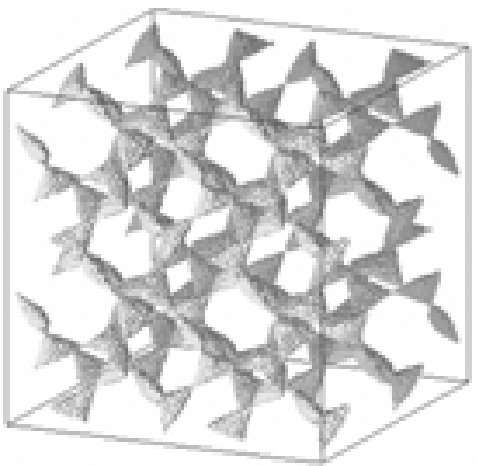}&\includegraphics[width=2.cm]{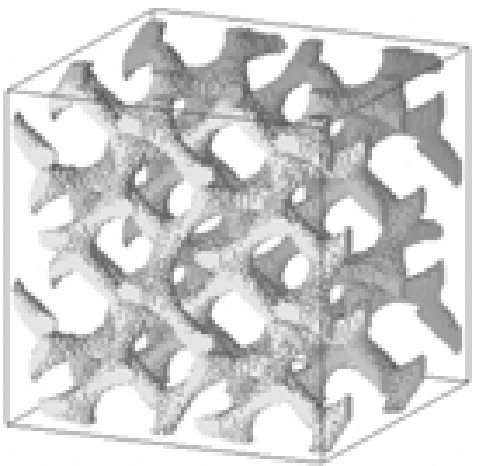}
&\includegraphics[width=2.cm]{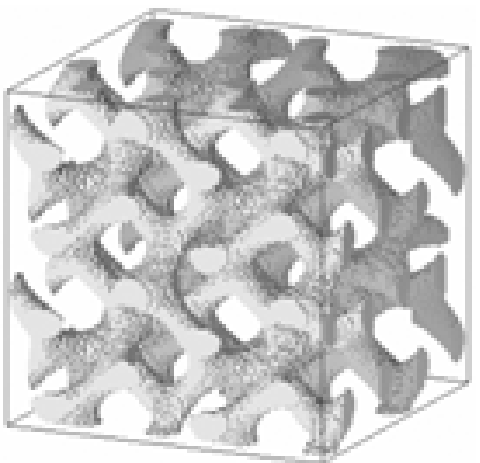}&\includegraphics[width=2.cm]{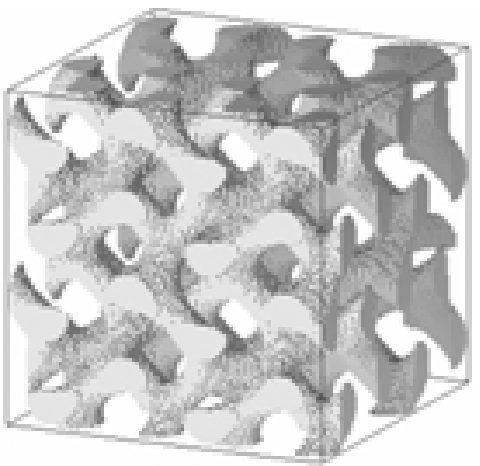}
&\includegraphics[width=2.cm]{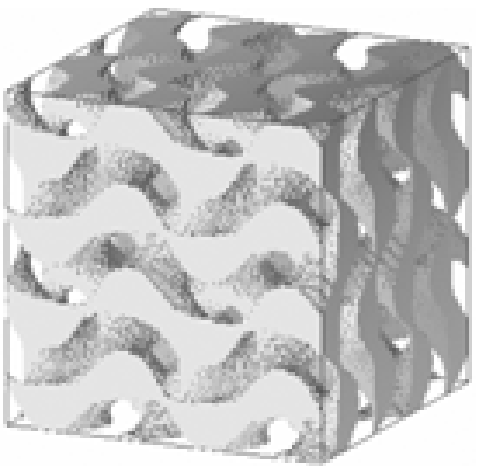}&\includegraphics[width=2.cm]{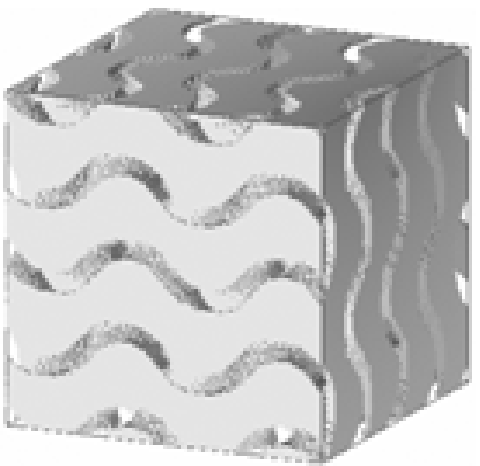}
& &\\
&$t=-5.0$&$t=-4.0$&$t=-3.0$&$t=-2.0$&$t=0.0$&$t=2.0$&
& &\\ 
&$\phi_s=0.03$&$\phi_s=0.10$&$\phi_s=0.18$&$\phi_s=0.26$&$\phi_s=0.43$&$\phi_s=0.68$&
& &\\ 
\noalign{\vskip 2pt\hrule\vskip 4pt}
inv I$_2$ - Y$^{**}$
&\multicolumn{9}{l}{
$-2[\sin 2X\cos Y\sin Z+\sin X\sin 2Y\cos Z+\cos X \sin Y\sin 2Z]$}\\
&\multicolumn{9}{l}{
$+\cos 2X\cos 2Y+\cos 2Y\cos 2Z+\cos 2X\cos 2Z=t$
}\\ \noalign{\vskip 4pt}
&\includegraphics[width=2.cm]{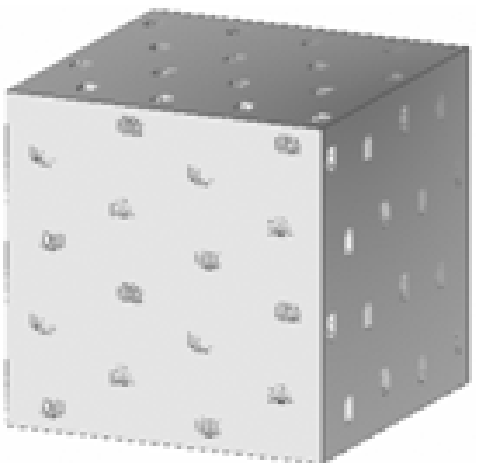}&\includegraphics[width=2.cm]{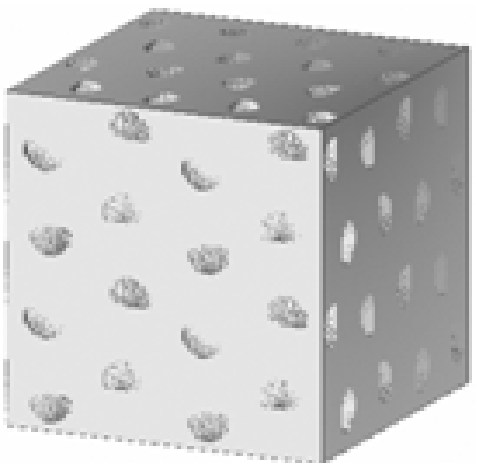}
&\includegraphics[width=2.cm]{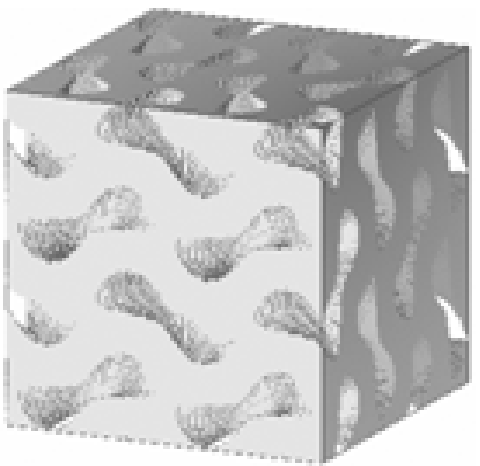}&\includegraphics[width=2.cm]{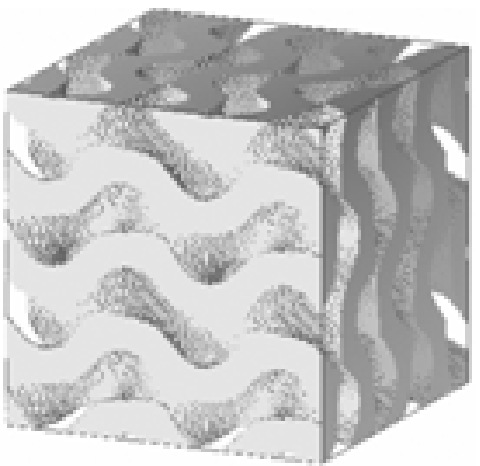}
&\includegraphics[width=2.cm]{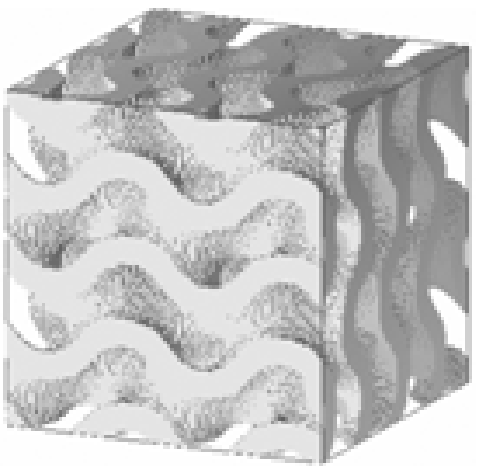}&\includegraphics[width=2.cm]{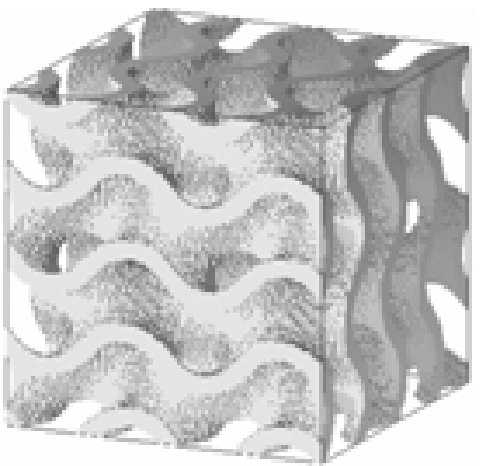}
&\includegraphics[width=2.cm]{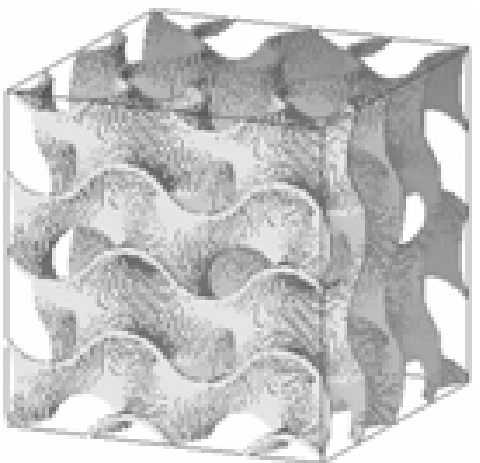}
& &\\
&$t=-4.0$&$t=-3.0$&$t=-1.0$&$t=0.0$&$t=1.0$&$t=2.0$&$t=2.9$
& &\\ 
&$\phi_s=0.90$&$\phi_s=0.82$&$\phi_s=0.66$&$\phi_s=0.57$&$\phi_s=0.46$&$\phi_s=0.323$&$\phi_s=0.10$
& &\\ 
\end{tabular}
\end{ruledtabular}
\end{center}
\end{table}

\begin{table}
\begin{center}
\caption{Same as Table 1}
\label{tab1c}
\begin{ruledtabular}
\begin{tabular}{ccccccccccc}
C(I$_2$ - Y$^{**}$)
&\multicolumn{9}{l}{ 
$2[\sin 2X\cos Y\sin Z+\sin X\sin 2Y\cos Z +\cos X\sin Y\sin 2Z]$}\\
&\multicolumn{9}{l}{
$+\cos 2X\cos 2Y+\cos 2Y\cos 2Z +\cos 2X\cos 2Z=t$
}\\ \noalign{\vskip 4pt}
&\includegraphics[width=2.cm]{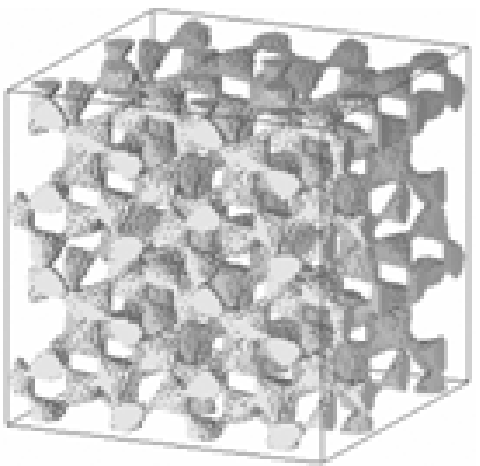}&\includegraphics[width=2.cm]{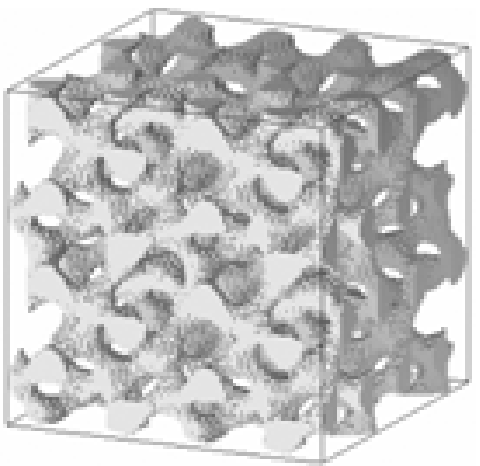}
&\includegraphics[width=2.cm]{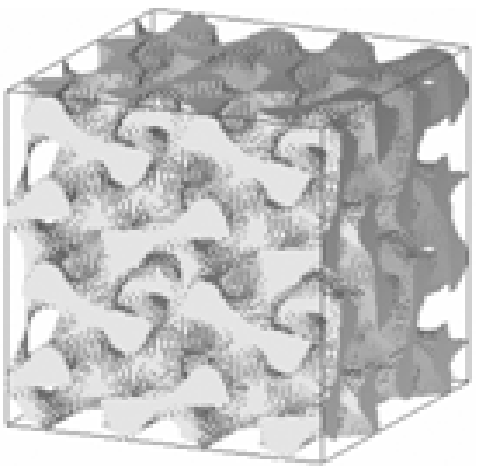}&\includegraphics[width=2.cm]{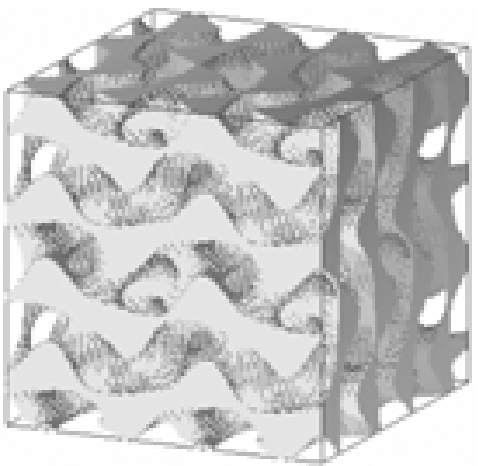}
&\includegraphics[width=2.cm]{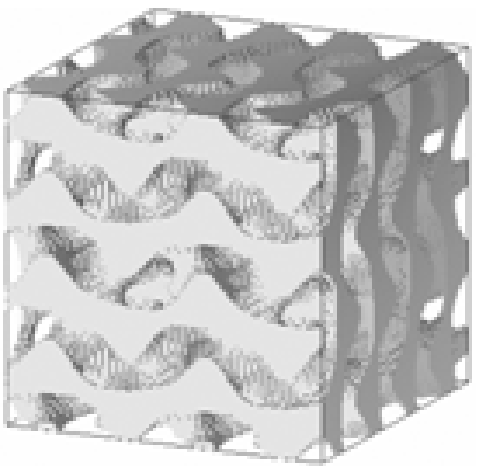}&\includegraphics[width=2.cm]{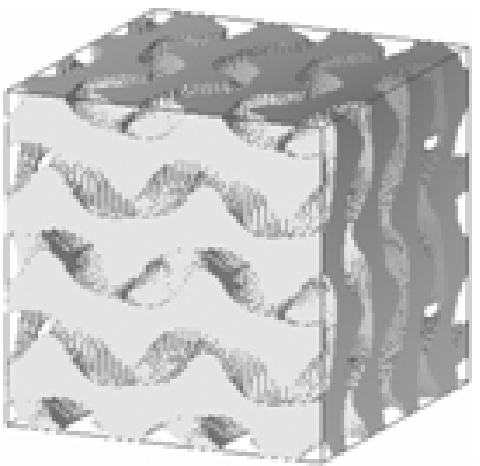}
&\includegraphics[width=2.cm]{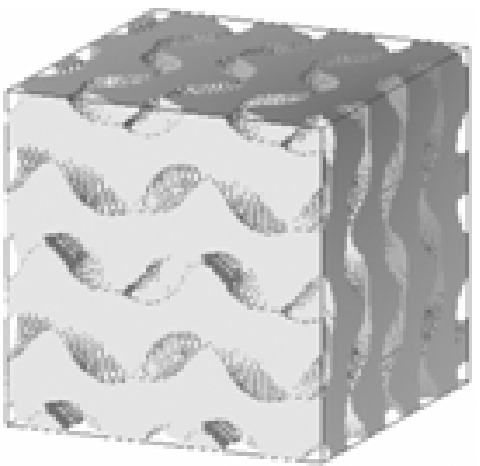}
& &\\
&$t=-2.0$&$t=-1.5$&$t=-1.0$&$t=-0.5$&$t=0.0$&$t=0.5$&$t=0.9$
& &\\ 
&$\phi_s=0.08$&$\phi_s=0.20$&$\phi_s=0.30$&$\phi_s=0.40$&$\phi_s=0.51$&$\phi_s=0.61$&$\phi_s=0.70$
& &\\ 
\noalign{\vskip 2pt\hrule\vskip 4pt}
inv C(I$_2$ - Y$^{**}$)
&\multicolumn{9}{l}{ 
$2[\sin 2X\cos Y\sin Z+\sin X\sin 2Y\cos Z +\cos X\sin Y\sin 2Z]$}\\
&\multicolumn{9}{l}{
$+\cos 2X\cos 2Y+\cos 2Y\cos 2Z +\cos 2X\cos 2Z=t$
}\\ \noalign{\vskip 4pt}
&\includegraphics[width=2.cm]{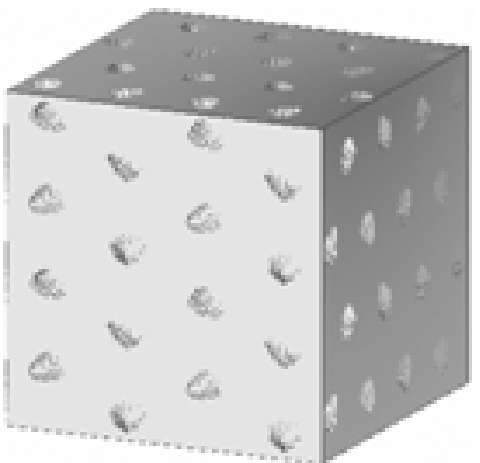}&\includegraphics[width=2.cm]{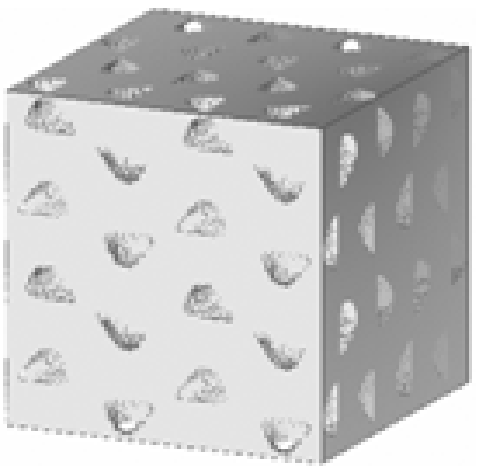}
&\includegraphics[width=2.cm]{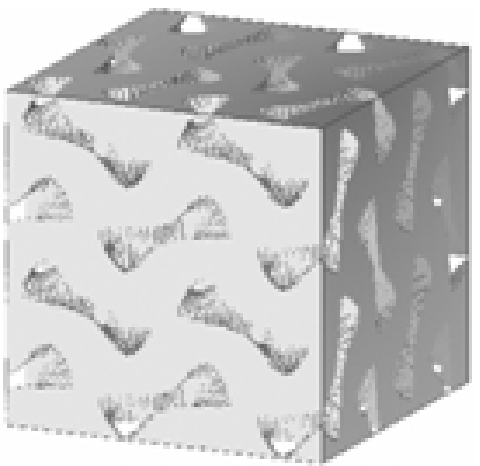}&\includegraphics[width=2.cm]{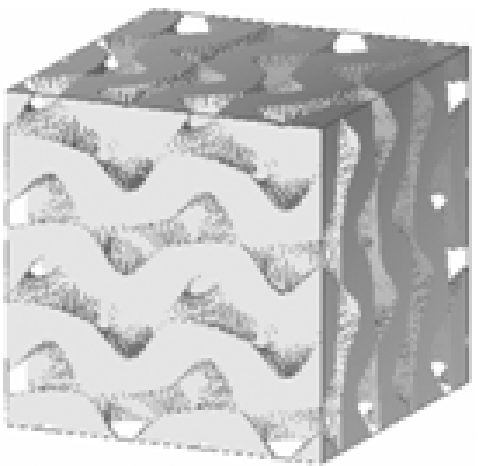}
&\includegraphics[width=2.cm]{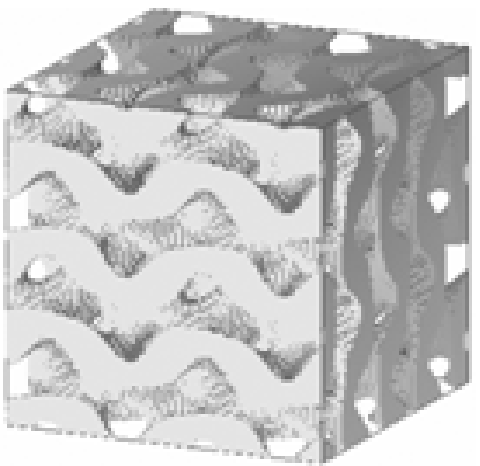}&\includegraphics[width=2.cm]{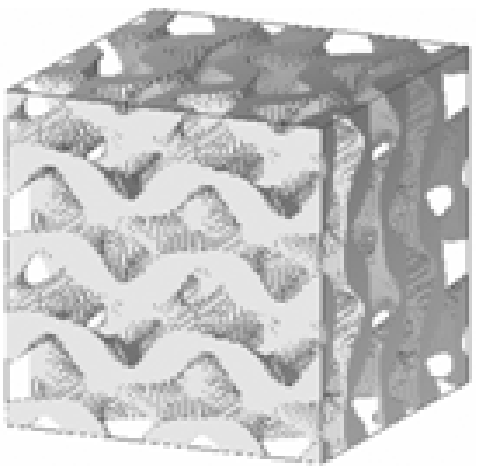}
&\includegraphics[width=2.cm]{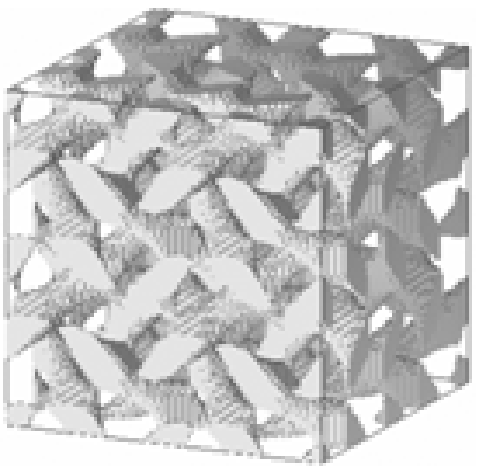}
& &\\
&$t=-2.0$&$t=-1.5$&$t=-1.0$&$t=-0.5$&$t=0.0$&$t=0.5$&$t=1.0$
& &\\
&$\phi_s=0.90$&$\phi_s=0.80$&$\phi_s=0.70$&$\phi_s=0.60$&$\phi_s=0.50$&$\phi_s=0.39$&$\phi_s=0.28$
& &\\
\noalign{\vskip 2pt\hrule\vskip 4pt}
L
&\multicolumn{9}{l}{ 
$0.5[\sin 2X\cos Y\sin Z+\sin 2Y\cos Z\sin X +\sin 2Z\cos X\sin Y] $}\\
&\multicolumn{9}{l}{
$-0.5[\cos 2X\cos 2Y +\cos 2Y\cos 2Z +\cos 2Z\cos 2X]+0.15=t$
}\\ \noalign{\vskip 4pt}
&\includegraphics[width=2.cm]{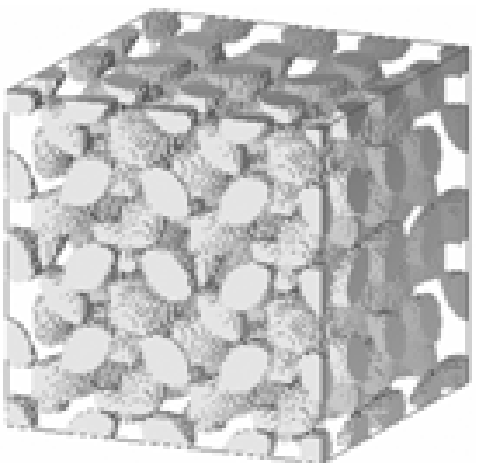}&\includegraphics[width=2.cm]{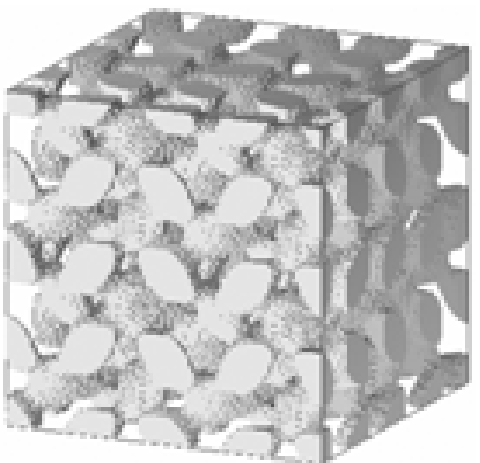}
&\includegraphics[width=2.cm]{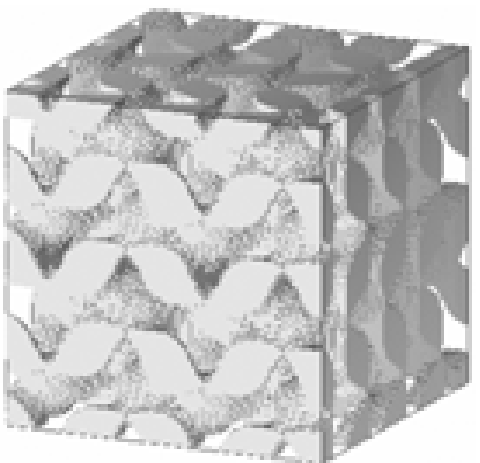}&\includegraphics[width=2.cm]{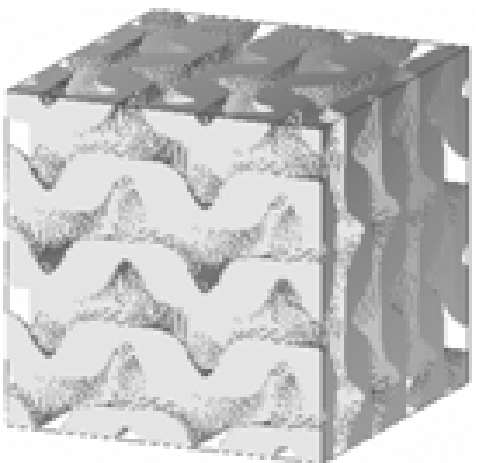}
&\includegraphics[width=2.cm]{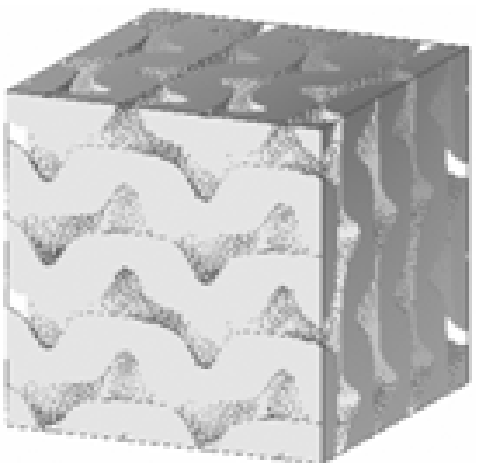}&\includegraphics[width=2.cm]{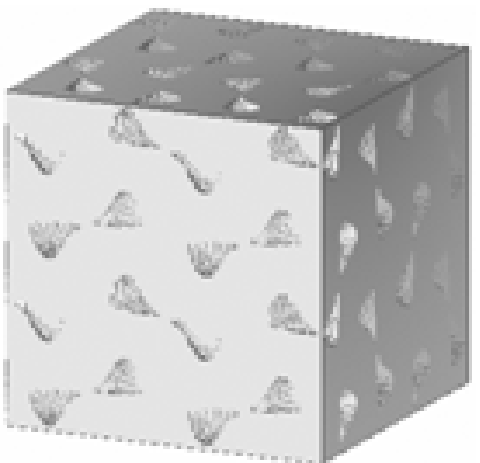}
& &\\
&$t=-0.3$&$t=0.0$&$t=0.3$&$t=0.5$&$t=1.0$&$t=1.5$&
& &\\ 
&$\phi_s=0.27$&$\phi_s=0.39$&$\phi_s=0.51$&$\phi_s=0.57$&$\phi_s=0.71$&$\phi_s=0.84$&
& &\\ 
\noalign{\vskip 2pt\hrule\vskip 4pt}
inv L
&\multicolumn{9}{l}{ 
$0.5[\sin 2X\cos Y\sin Z+\sin 2Y\cos Z\sin x +\sin 2Z\cos X\sin Y]$}\\
&\multicolumn{9}{l}{
$ -0.5[\cos 2X\cos 2Y +\cos 2Y\cos 2Z +\cos 2Z\cos 2X]+0.15=t$
}\\ \noalign{\vskip 4pt}
&\includegraphics[width=2.cm]{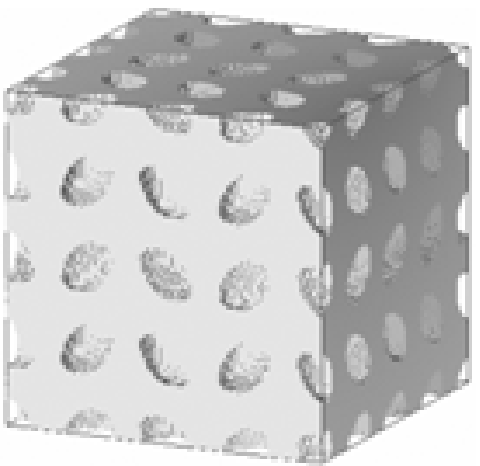}&\includegraphics[width=2.cm]{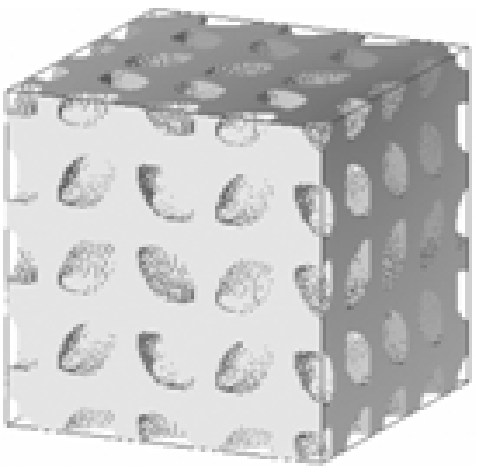}
&\includegraphics[width=2.cm]{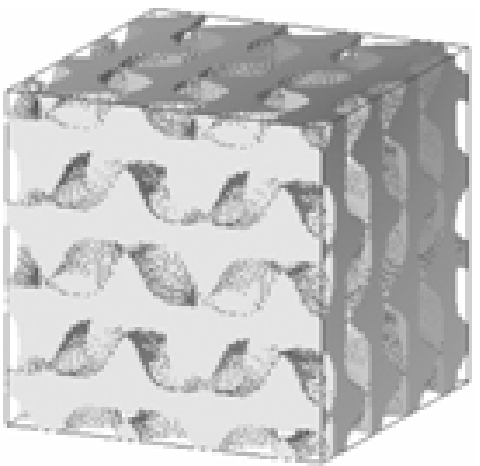}&\includegraphics[width=2.cm]{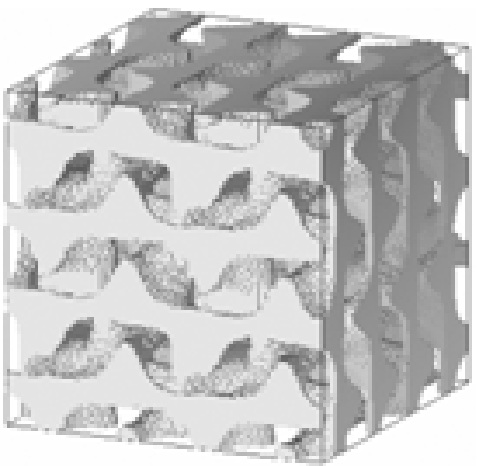}
&\includegraphics[width=2.cm]{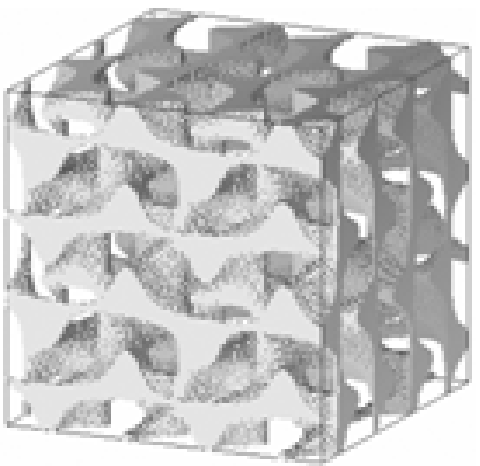}&\includegraphics[width=2.cm]{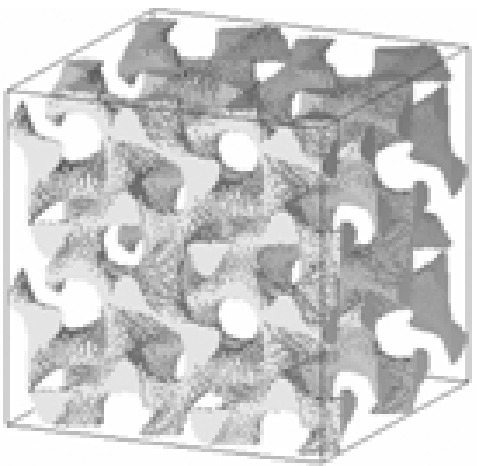}
& &\\
&$t=-0.3$&$t=0.0$&$t=0.3$&$t=0.5$&$t=1.0$&$t=1.5$&
& &\\ 
&$\phi_s=0.73$&$\phi_s=0.62$&$\phi_s=0.50$&$\phi_s=0.43$&$\phi_s=0.29$&$\phi_s=0.16$&
& &\\ 
\end{tabular}
\end{ruledtabular}
\end{center}
\end{table}

\begin{table}
\begin{center}
\caption{Maximum band gap widths $\Delta\omega /\omega_0$, corresponding
midgap frequency $\omega_0$, volume fraction $\phi_s$ and parameter $t$
of various single level surface structures,
categorized by their space group and genus $g$.
A dash indicates that the structure has no photonic band gap.
$^a$ The value found in the literature for the genus for the crystallographic cell of the minimal C(S) surface
is $g=17$~\cite{Fisher}.
$^b$ The value found in the literature for the genus for the crystallographic cell of the minimal Y surface
is $g=17$~\cite{Fisher}.
$^c$ Same value as reported by
Garstecki {\sl et al.}~\cite{Garstecki2}, but different from the value $g=145$ reported
by Fisher {\sl et al.}~\cite{Fisher} for the minimal D surface.
For an explanation of the deviating values of the genus see text.}
\label{tbl:minsurf2}
\begin{ruledtabular}
\begin{tabular}{ccccccc}
Space group (index) & Surface & $g$ & $\phi_s$ (\%) & $t$ & $\Delta\omega /\omega_0$ (\%) & $\omega_0$\\
\hline
$Pa{\bar 3}$ (205) & C($^{\pm}$Y) & 13 & 17 & -1.30 & 8.1 & 0.786\\
& $^{\pm}$Y & 21 & - & - & - & -\\
\hline
$Ia{\bar 3}$ (206) & C(S) & 65$^a$ & - & - & - & -\\
\hline
$P4_332$ (212) & C(Y) & 13 & 22 & -1.25 & 6.2 & 0.630\\
& Y & 13$^b$ & - & - & - & -\\
\hline
$I4_132$ (214) & G & 5 & 21 & -0.90 & 21.7 & 0.474\\
 & C(Y$^{**}$) & 29& 18 & -3.25 & 11.9 & 0.514\\
\hline
$I{\bar 4}3d$ (220) & S & 21 & - & - & - & -\\
\hline
$Pm{\bar 3}m$ (221) & P & 3 & 24 & -0.90 & 6.8 & 0.442\\
 & I-WP & 7 & - & - & - & -\\
 & inv I-WP & 7 & - & - & - & -\\
 & C(P) & 9 & - & - & - & -\\
\hline
$Fm{\bar 3}m$ (225) & F-RD & 21 & 31 & -0.60 & 6.6 & 0.773\\
 & inv F-RD & 21 & 39 & 0.80 & 4.2 & 0.858\\
\hline
$Fd{\bar 3}m$ (227) & D & 9 & 25 & -0.60 & 20.3 & 0.541\\
& C(D) & 121$^c$ & - & - & - & -\\
\hline
$Ia{\bar 3}d$ (230) & I$_2$-Y$^{**}$ & 9 & - & - & - & -\\
 & inv I$_2$-Y$^{**}$ & 9 & - & - & - & -\\
 & C(I$_2$-Y$^{**}$) & 25 & 25 & -1.25 & 19.6 & 0.796\\
 & inv C(I$_2$-Y$^{**}$) & 25 & 39 & 0.50 & 4.7 & 0.675\\
 & L & 33 & - & - & - & -\\
 & inv L & 33 & - & - & - & -\\
\end{tabular}
\end{ruledtabular}
\end{center}
\end{table}

\begin{table}
\begin{center}
\caption{Equations and 3D renderings of 2x2x2 unit cubes of various tubular structures for various values of $t$
and corresponding volume fractions $\phi_s$.
$M=2\pi m/a$ and $M^{\prime}=M-\pi /4$, where $M=X,Y,Z$, $m=x,y,z$ and $a$ denotes
the length of the crystallographic cell.}
\label{tab1d}
\begin{ruledtabular}
\begin{tabular}{ccccccccccc}
tubular P
&\multicolumn{9}{l}{ 
$10[\cos X +\cos Y +\cos Z]-5.1[\cos X\cos Y +\cos Y\cos Z +\cos Z\cos X]-14.6=t$
}\\ \noalign{\vskip 4pt}
&\includegraphics[width=2.cm]{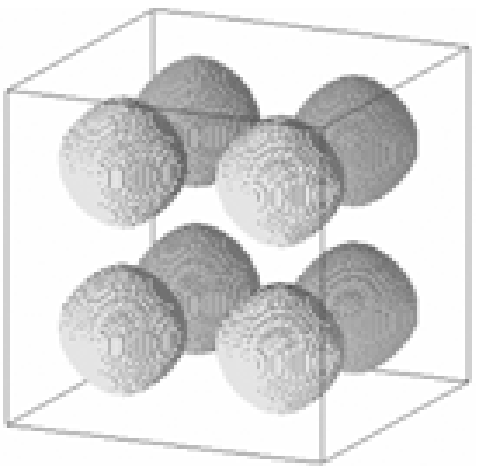}&\includegraphics[width=2.cm]{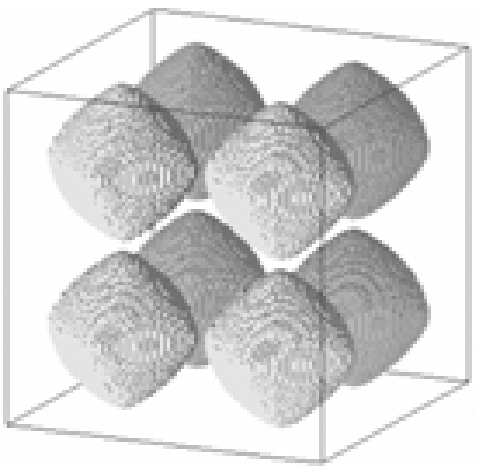}
&\includegraphics[width=2.cm]{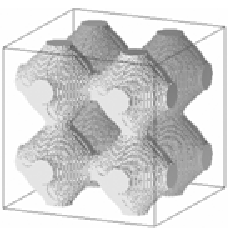}&\includegraphics[width=2.cm]{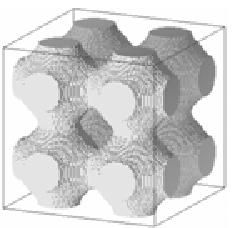}
&\includegraphics[width=2.cm]{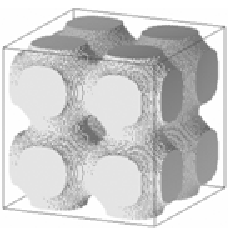}&\includegraphics[width=2.cm]{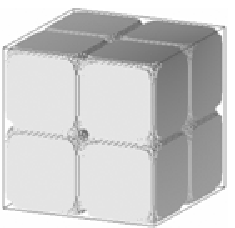}
& &
& &\\
&$t=-25.0$&$t=-20.0$&$t=-15.0$&$t=-10.0$&$t=-5.0$&$t=0.0$&
& &\\
&$\phi_s=0.19$&$\phi_s=0.27$&$\phi_s=0.41$&$\phi_s=0.54$&$\phi_s=0.71$&$\phi_s=0.97$&
& &\\
\noalign{\vskip 2pt\hrule\vskip 4pt}
inv tubular P
&\multicolumn{9}{l}{ 
$10[\cos X +\cos Y +\cos Z]-5.1[\cos X\cos Y +\cos Y\cos Z +\cos Z\cos X]-14.6=t$
}\\ \noalign{\vskip 4pt}
&\includegraphics[width=2.cm]{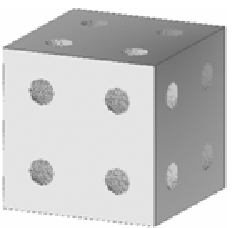}&\includegraphics[width=2.cm]{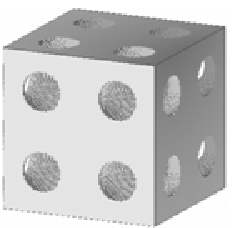}
&\includegraphics[width=2.cm]{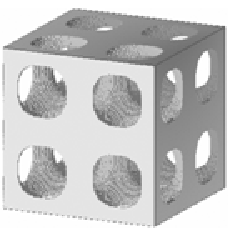}&\includegraphics[width=2.cm]{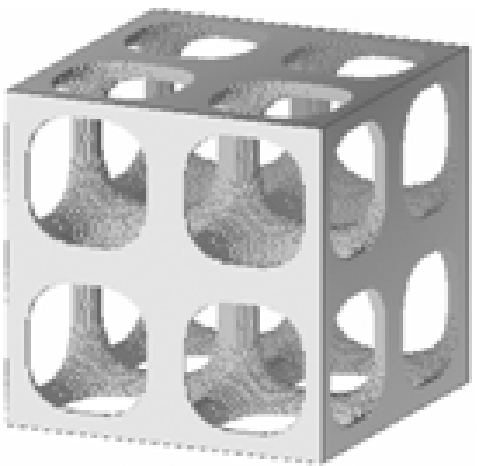}
&\includegraphics[width=2.cm]{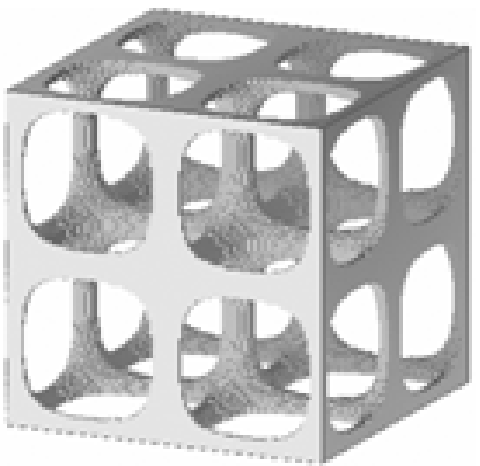}&\includegraphics[width=2.cm]{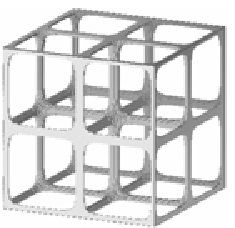}
& %
& &\\ 
&$t=-15.0$&$t=-10.0$&$t=-5.0$&$t=-2.0$&$t=-1.0$&$t=0.0$&
& &\\ 
&$\phi_s=0.60$&$\phi_s=0.46$&$\phi_s=0.29$&$\phi_s=0.16$&$\phi_s=0.11$&$\phi_s=0.03$&
& &\\ 
\noalign{\vskip 2pt\hrule\vskip 4pt}
tubular D
&\multicolumn{9}{l}{ 
$10[\sin X^{\prime}\sin Y^{\prime}\sin Z^{\prime} +\sin X^{\prime}\cos Y^{\prime}
\cos Z^{\prime} + \cos X^{\prime}\sin Y^{\prime}\cos Z^{\prime} +
\cos X^{\prime}\cos Y^{\prime}\sin Z^{\prime}]$}\\
&\multicolumn{9}{l}{
$-0.7[\cos 4X+\cos 4Y+\cos 4Z]-11=t$
}\\ \noalign{\vskip 4pt}
&\includegraphics[width=2.cm]{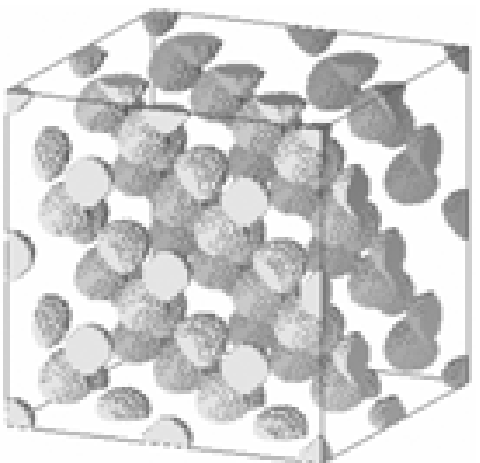}&\includegraphics[width=2.cm]{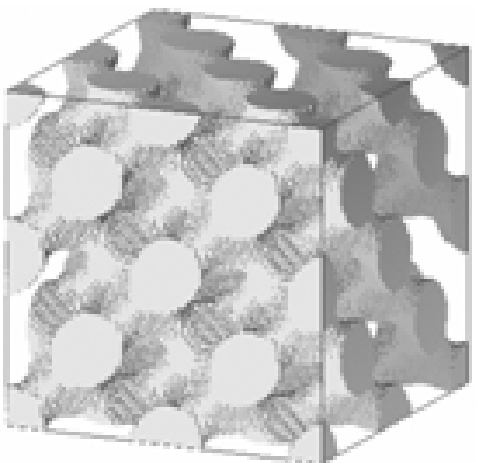}
&\includegraphics[width=2.cm]{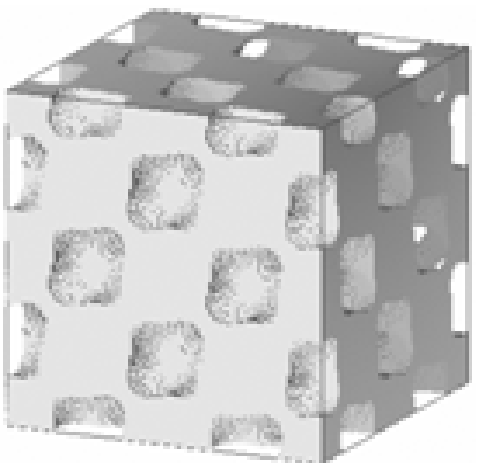}&\includegraphics[width=2.cm]{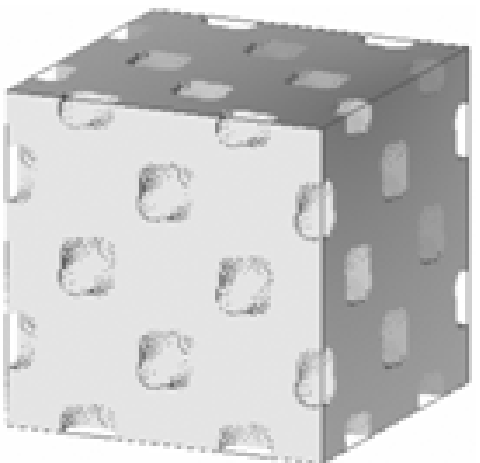}
&\includegraphics[width=2.cm]{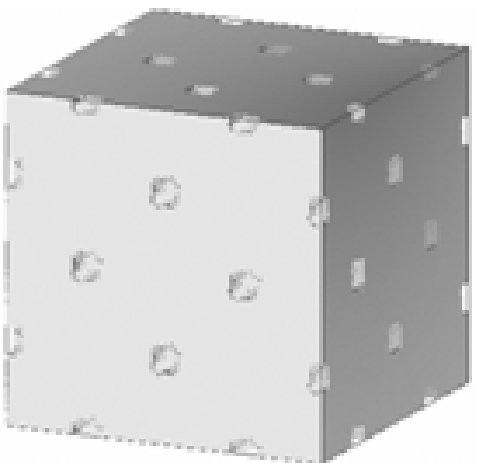}&\includegraphics[width=2.cm]{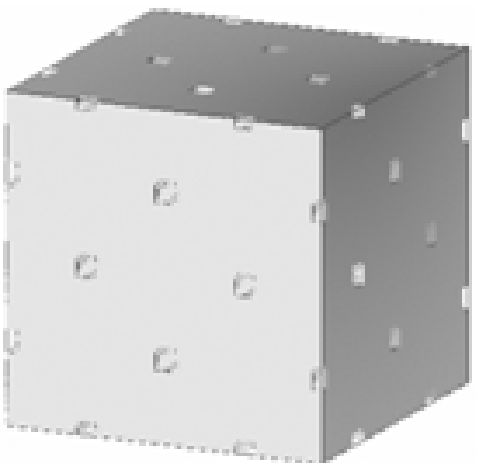}
& &\\
&$t=-20.0$&$t=-15.0$&$t=-10.0$&$t=-5.0$&$t=0.0$&$t=0.5$&
& &\\
&$\phi_s=0.10$&$\phi_s=0.35$&$\phi_s=0.55$&$\phi_s=0.74$&$\phi_s=0.94$&$\phi_s=0.97$&
& &\\
\noalign{\vskip 2pt\hrule\vskip 4pt}
inv tubular D
&\multicolumn{9}{l}{ 
$10[\sin X^{\prime}\sin Y^{\prime}\sin Z^{\prime} +\sin X^{\prime}\cos Y^{\prime}
\cos Z^{\prime} + \cos X^{\prime}\sin Y^{\prime}\cos Z^{\prime} +
\cos X^{\prime}\cos Y^{\prime}\sin Z^{\prime}]$}\\
&\multicolumn{9}{l}{
$-0.7[\cos 4X+\cos 4Y+\cos 4Z]-11=t$
}\\ \noalign{\vskip 4pt}
&\includegraphics[width=2.cm]{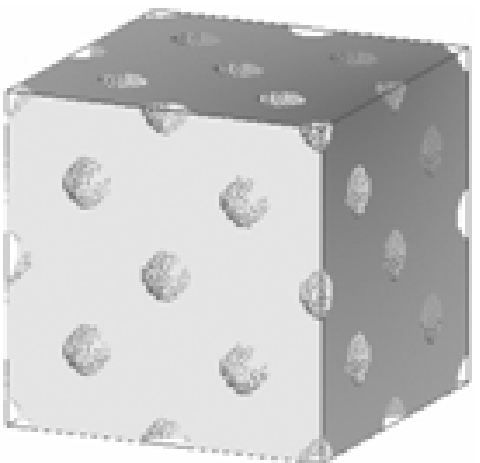}&\includegraphics[width=2.cm]{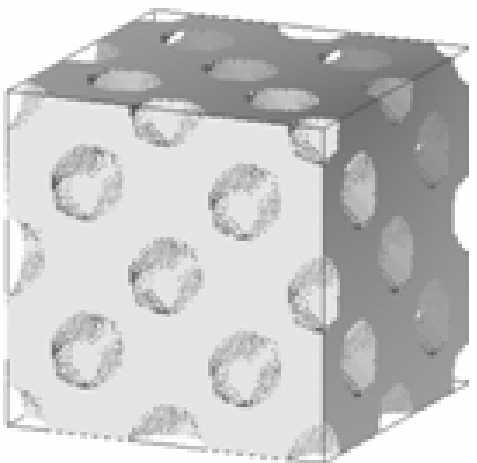}
&\includegraphics[width=2.cm]{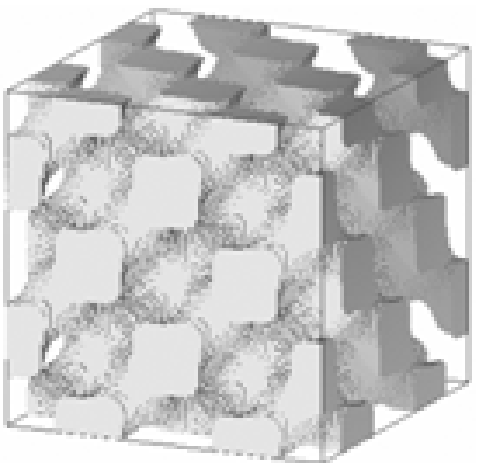}&\includegraphics[width=2.cm]{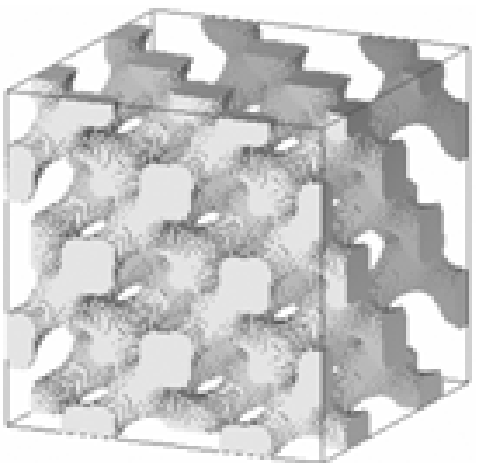}
&\includegraphics[width=2.cm]{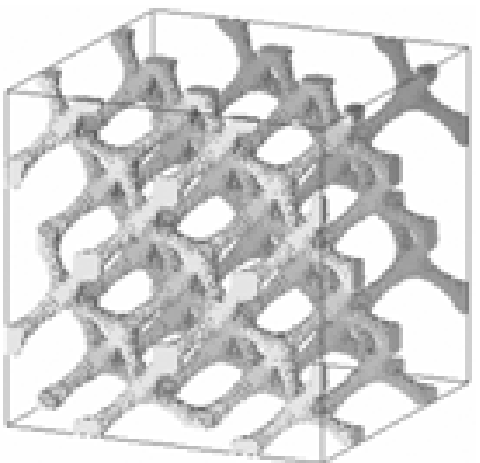}&\includegraphics[width=2.cm]{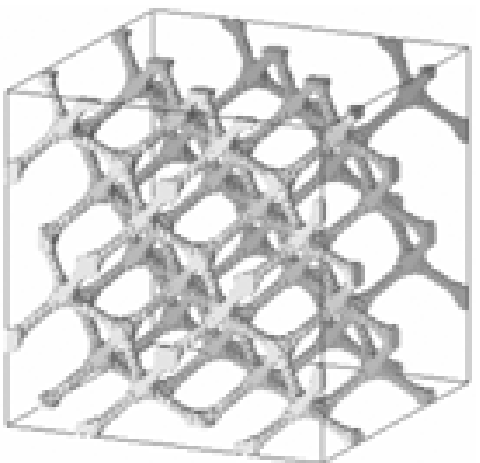}
& %
& &\\ 
&$t=-20.0$&$t=-15.0$&$t=-10.0$&$t=-5.0$&$t=0.0$&$t=0.5$&
& &\\ 
&$\phi_s=0.99$&$\phi_s=0.65$&$\phi_s=0.45$&$\phi_s=0.26$&$\phi_s=0.06$&$\phi_s=0.03$&
& &\\
\noalign{\vskip 2pt\hrule\vskip 4pt}
tubular G
&\multicolumn{9}{l}{ 
$10[\cos X\sin Y +\cos Y\sin Z +\cos Z\sin X]-0.5[\cos 2X\cos 2Y +
\cos 2Y\cos 2Z +\cos 2Z\cos 2X]-14=t$
}\\ \noalign{\vskip 4pt}
&\includegraphics[width=2.cm]{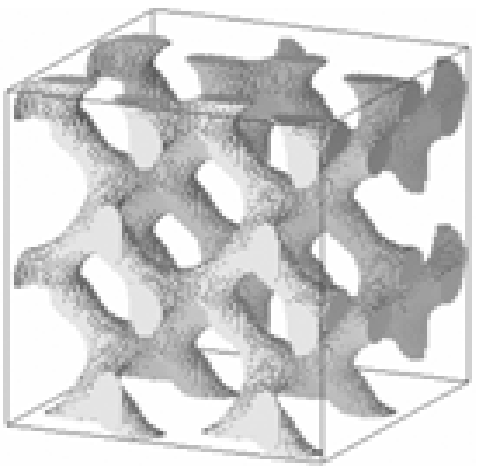}&\includegraphics[width=2.cm]{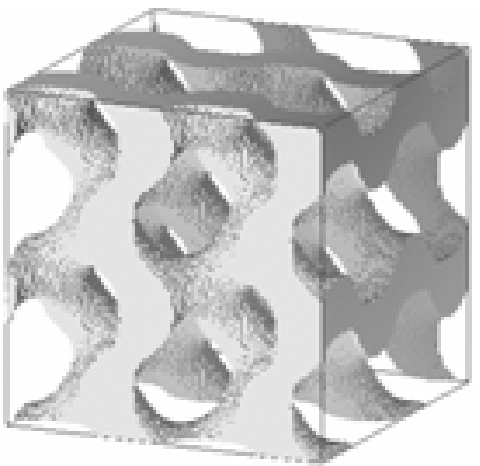}
&\includegraphics[width=2.cm]{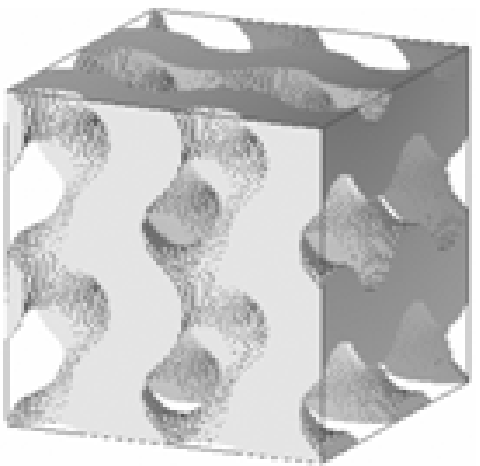}&\includegraphics[width=2.cm]{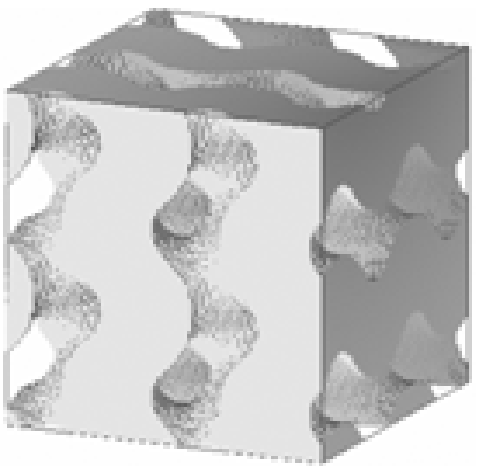}
&\includegraphics[width=2.cm]{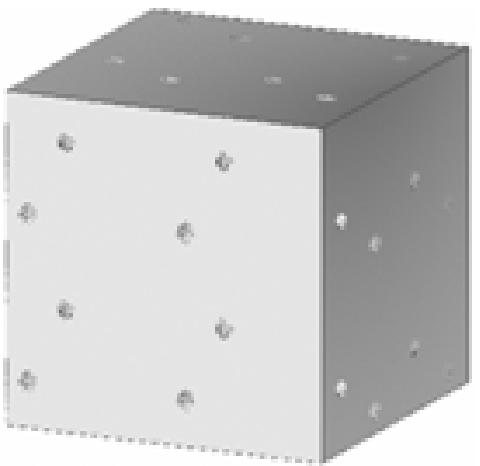}&\includegraphics[width=2.cm]{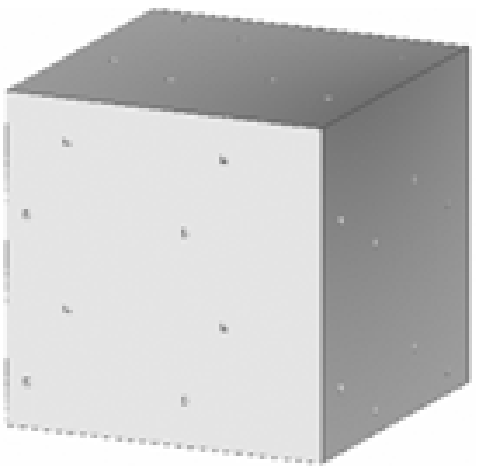}& %
& %
& &\\ 
&$t=-25.0$&$t=-20.0$&$t=-15.0$&$t=-10.0$&$t=0.0$&$t=0.5$&
& &\\ 
&$\phi_s=0.13$&$\phi_s=0.31$&$\phi_s=0.47$&$\phi_s=0.63$&$\phi_s=0.97$&$\phi_s=0.99$& &
& &\\ 
\noalign{\vskip 2pt\hrule\vskip 4pt}
inv tubular G
&\multicolumn{9}{l}{ 
$10[\cos X\sin Y +\cos Y\sin Z +\cos Z\sin X]-0.5[\cos 2X\cos 2Y +
\cos 2Y\cos 2Z +\cos 2Z\cos 2X]-14=t$
}\\ \noalign{\vskip 4pt}
&\includegraphics[width=2.cm]{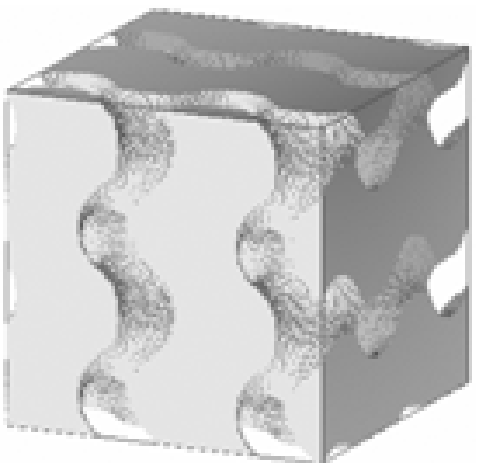}&\includegraphics[width=2.cm]{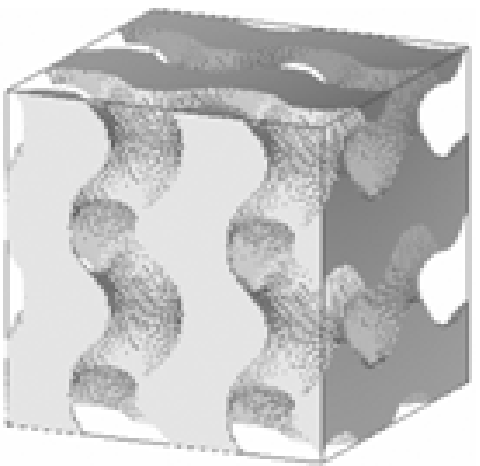}
&\includegraphics[width=2.cm]{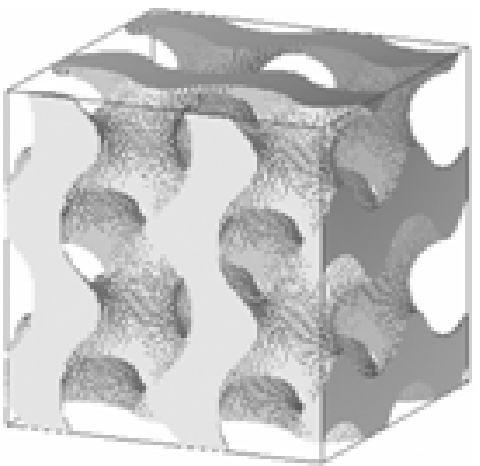}&\includegraphics[width=2.cm]{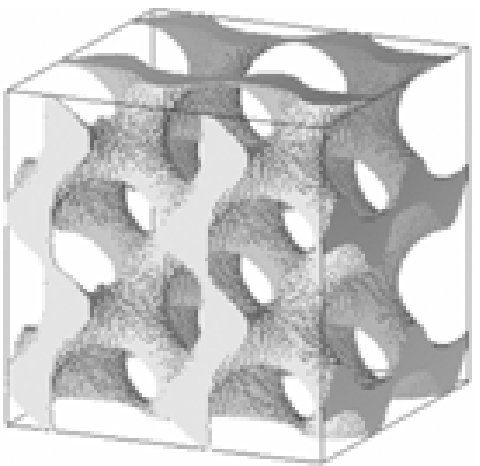}
&\includegraphics[width=2.cm]{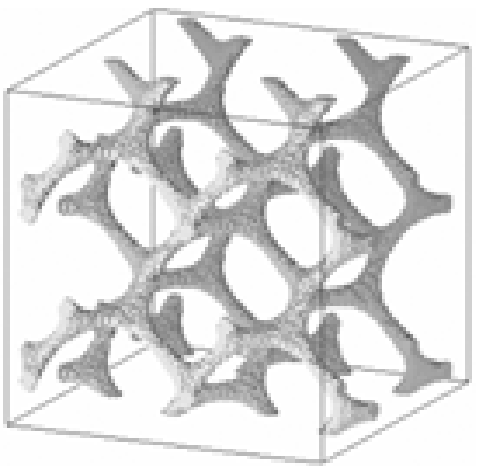}&\includegraphics[width=2.cm]{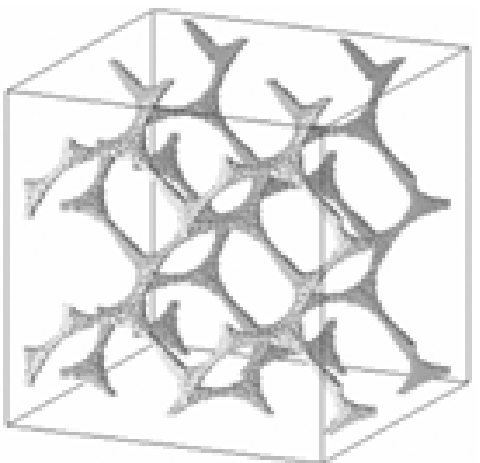}
& %
& &\\ 
&$t=-20.0$&$t=-15.0$&$t=-10.0$&$t=-5.0$&$t=0.0$&$t=0.5$&
& &\\ 
&$\phi_s=0.69$&$\phi_s=0.53$&$\phi_s=0.37$&$\phi_s=0.21$&$\phi_s=0.03$&$\phi_s=0.01$&
& &\\ 
\end{tabular}
\end{ruledtabular}
\end{center}
\end{table}

\begin{table}
\begin{center}
\caption{Maximum band gap widths $\Delta\omega /\omega_0$, corresponding
midgap frequency $\omega_0$, volume fraction $\phi_s$ and parameter $t$
of various tubular structures, categorized by their genus $g$.
A dash indicates that the structure has no photonic band gap.}
\label{tbl:minsurf3}
\begin{ruledtabular}
\begin{tabular}{cccccc}
Tubular structure & $g$ & $\phi_s$ (\%) & $t$ & $\Delta\omega /\omega_0$ (\%) & $\omega_0$\\
\hline
P & 3 & 27 & -16.50 & 4.6 & 0.427\\
inv P & 3 & - & - & - & -\\
\hline
G & 5 & 17 & -24.00 & 20.4 & 0.514\\
inv G & 5 & 17 & -4.00 & 19.9 & 0.502\\
\hline
D & 9 & 25 & -17.00 & 19.2 & 0.546\\
inv D & 9 & 22 & -4.00 & 21.4 & 0.569\\
\end{tabular}
\end{ruledtabular}
\end{center}
\end{table}

\end{document}